\documentclass[reprint,
 amsmath,amssymb,
aps,
pra,
]{revtex4-2}
\usepackage{mathrsfs}
\usepackage{caption}
\usepackage{graphicx}
\usepackage{dcolumn}
\usepackage{bm}
\usepackage{float}
\usepackage{subfig}
\usepackage{color}
\usepackage{hyperref}
\begin{document}

\preprint{APS/123-QED}

\title{Nonclassical Photon-Bundle Correlations in Quantum Rabi Models}



	

\author{Yong-Xin Zhang$^{1}$}
\author{Chen Wang$^{2,}$}
\email{wangchen@zjnu.cn}
\author{Qing-Hu Chen$^{1,3,}$}
\email{qhchen@zju.edu.cn}
\affiliation{$^{1}$Zhejiang Key Laboratory of Micro-Nano Quantum Chips and Quantum Control, School of Physics, Zhejiang University, Hangzhou 310027, China\\
                $^{2}$Department of Physics, Zhejiang Normal University, Jinhua 321004, China\\
		$^{3}$Collaborative Innovation Center of Advanced Microstructures, Nanjing University, Nanjing 210093, China}

\date{\today}

\begin{abstract} 
We investigate nonclassical photon-bundle correlations in the quantum Rabi model and its extended cases, using the quantum dressed master equation. By tuning the light–matter coupling strength at finite temperature, the quantum Rabi model exhibits controllable nonclassical transitions between two-photon bundle bunching and antibunching, allowing for the two-photon bundle emission and statistics. We further introduce anisotropic coupling and nonlinear Stark interactions, which enrich the photon statistical behaviors and provide additional tunability of photon-bundle correlations. Extreme correlation behaviors are found to be closely linked to excited-state quantum phase transitions, suggesting a potential pathway for predicting and exploiting excited-state phenomena. These effects can be controlled solely by tuning intrinsic system parameters, without the need for an external modulating field. The quantum Rabi model family thus provides a flexible and experimentally feasible platform for high-purity photon bundle generation and controllable multi-photon sources.
\end{abstract}


\maketitle

\section{Introduction}
Nonclassical light is the cornerstone of modern quantum optics, revealing phenomena beyond the reach of classical electromagnetic theory~\cite{Scully1997__}. These genuinely quantum effects constitute a central subject in modern quantum optics~\cite{1998__,Haroche2006__,Carmichael2008__}. Since the pioneering experiment of Brown and Twiss in 1956~\cite{BROWN1956N_177_27-29}, the study of quantum coherence and photon correlations has provided a fundamental framework for identifying and characterizing quantum states of light~\cite{Glauber1963PR_130_2529-2539}. The observation of classically forbidden effects not only reveals the quantum nature of light, but also spurs  the development of emerging quantum technologies, such as quantum computation~\cite{Kok2007RMP_79_135-174,Raussendorf2007PRL_98_190504,Aaronson2011PFAASTC__333–342,Gambetta2017nQI_3_2,Liu2019NN_14_586-593,Wang2019NP_13_770-775}, secure communication~\cite{Cleve1997PRA_56_1201-1204,Braunstein2005RMP_77_513-577,Kimble2008N_453_1023-1030,Scarani2009RMP_81_1301-1350,Wang2019PRA_99_042309}, and high-precision measurement~\cite{Caves1980RMP_52_341-392b,Grangier1987PRL_59_2153-2156,Xiao1987PRL_59_278-281,Polzik1992PRL_68_3020-3023,Aasi2013NP_7_613-619,Pedrozo-Penafiel2020N_588_414-418,LIGOScientificCollaborationandVirgoCollaboration2021PRX_11_021053,Grote2013PRL_110_181101}.

Controlling the quantum character of light is essential for photonic quantum technologies, especially with regard to photon correlation statistics~\cite{Glauber1963PR_130_2529-2539}. A paradigmatic example is photon blockade, characterized by $G^{(2)}(0)<1$, which gives rise to temporal antibunching between individual photons and enables high-quality single-photon sources~\cite{Scully1997__,Flayac2017PRA_96_053810}. However, in systems subject to the strong-coupling or strong-nonlinearity regimes, single-photon transitions can be strongly suppressed~\cite{Imamoglu1997PRL_79_1467-1470,Bienfait2019S__}. Photon emission may then occur in the form of correlated groups of photons, known as photon bundles~\cite{Munoz2014NP_8_550-555,Strekalov2014NP_8_500-501}. In such situations, the conventional single-photon description becomes inadequate~\cite{Scully1997__,Ridolfo2012PRL_109_193602,Ridolfo2013PRL_110_163601}, and the system behavior must instead be characterized by correlations between photon bundles~\cite{Munoz2014NP_8_550-555,Strekalov2014NP_8_500-501}, where each bundle contains multiple photons. This bundle-based description has direct relevance to applications in quantum metrology~\cite{Motes2015PRL_114_170802,Moller2017N_547_191-195,Colombo2022NP_18_925-930,Qin2023PRL_130_070801}, lithography~\cite{Boto2000PRL_85_2733-2736,dangeloTwoPhotonDiffractionQuantum2001}, spectroscopy~\cite{lopezcarrenoExcitingPolaritonsQuantum2015,dorfmanNonlinearOpticalSignals2016}, and biological detection~\cite{denkTwoPhotonLaserScanning1990a,hortonVivoThreephotonMicroscopy2013,liTwophotonNanoprobesBased2023}.

The quantum Rabi model (QRM)~\cite{Rabi1936PR_49_324-328,Braak2016JPAMT_49_300301} describes the fundamental interaction between a two-level system and a single-mode quantized electromagnetic field. Despite its conceptual simplicity, QRM exhibits many nonclassical phenomena, including anomalous photon blockade~\cite{Ridolfo2013PRL_110_163601,Chen2022JPBAMOP_55_115502,chen2022prr}, and quantum squeezing~\cite{Genoni2015NJP_17_013034,Chen2022JPBAMOP_55_115502}. Its broad applicability has been demonstrated across a variety of platforms, such as superconducting circuit QED  systems~\cite{niemczykCircuitQuantumElectrodynamics2010,Yoshihara2017NP_13_44-47,forn-diazUltrastrongCouplingRegimes2019}, waveguide QED implementations~\cite{forn-diazUltrastrongCouplingSingle2017}, trapped-ion implementations~\cite{leibfriedQuantumDynamicsSingle2003,lvQuantumSimulationQuantum2018}, and semiconductor quantum dots--cavity platforms~\cite{englundControllingCavityReflectivity2007,hennessyQuantumNatureStrongly2007}. 
 When the rotating-wave (RW) and counter-rotating-wave (CRW) terms have unequal coupling strengths, the QRM generalizes to the anisotropic quantum Rabi model (AQRM), which displays first-order quantum phase transitions (QPTs)~\cite{Xie2014PRX_4_021046,Liu2017PRL_119_,Chen2021PRA_103_043708} and enhanced photon antibunching at finite temperatures~\cite{Ye2023PSMIA_609_128364,Ye2024OE_32_33483}. Moreover, incorporating nonlinear Stark coupling leads to the quantum Rabi–Stark model (QRSM)~\cite{Grimsmo2013PRA_87_033814,Grimsmo2014PRA_89_} and its anisotropic extension (AQRSM)~\cite{Zhang2018PRA_97_043858}, which exhibit rich phenomena such as spectral collapse, first-order and continuous QPTs~\cite{Eckle2017JPAMT_50_294004,Xie2019JPAMT_52_245304,Chen2020PRA_102_,xieFirstorderContinuousQuantum2020,Ying2023AQT_6_2200068,Braak2024JOSAB_41_C97}, thus offering enhanced controllability of quantum correlations. In particular, recent advances have also shown that AQRSM-based systems  not only outperform  conventional harmonic-oscillator platforms in quantum thermal machines~\cite{xuExploringRoleCriticality2024}, but also  enable superior control over nonclassical photon statistics and reduced field noise~\cite{zhangNonclassicalCorrelationsQuadrature}. In parallel, theoretical efforts have extended these models to open quantum systems by constructing non-Hermitian effective Hamiltonians~\cite{Tian2023JPBAMOP_56_095001a,Lu2023PRA_108_053712,Ying2024AQT_7_2400288,Li2025AQT_8_2400609,Wang2025PRA_112_043704,Wang2025AQT_n/a_e00622,jiang2025cpl}. These works thereby establish a distinct paradigm of nonclassical physics, centrally characterized by the emergence of spectral singularities, most notably exceptional points, and their associated dynamical signatures.

Until now, various schemes for generating multi-photon states and photon bundles have been proposed, ranging from photonic waveguides~\cite{liaoCorrelatedTwophotonTransport2010,douglasPhotonMoleculesAtomic2016,gonzalez-tudelaEfficientMultiphotonGeneration2017a,xingDeterministicGenerationArbitrary2024a}, Rydberg atomic ensembles~\cite{bieniasScatteringResonancesBound2014,jachymskiThreeBodyInteractionRydberg2016,liangObservationThreephotonBound2018}, cavity optomechanical platforms~\cite{liaoCorrelatedTwophotonScattering2013,Jing2014PRL_113_053604,Qin2019PRA_100_062501,zouDynamicalEmissionPhonon2022b,Bin2024PRL_133_043601}, to cavity QED (cQED) systems~\cite{Munoz2014NP_8_550-555,Strekalov2014NP_8_500-501,Chang2016PRL_117_203602a,Munoz2018OO_5_14-26,Bin2018PRA_98_043858a,Bin2020PRL_124_053601,Bin2021PRL_127_073602,Jiang2023OE_31_15697,Xiong2025PRR_7_013238}. Among these, cQED systems have attracted particular attention. By introducing external driving fields~\cite{Bin2018PRA_98_043858a,Bin2020PRL_124_053601,Bin2021PRL_127_073602},  multiphoton coupling schemes~\cite{Jiang2023OE_31_15697}, or engineered dissipation channels~\cite{Xiong2025PRR_7_013238}, multiphoton emission has been demonstrated theoretically. Nevertheless, many existing approaches rely on weak-coupling approximations~\cite{Munoz2014NP_8_550-555,Strekalov2014NP_8_500-501,Chang2016PRL_117_203602a,Munoz2018OO_5_14-26,Bin2018PRA_98_043858a,Jiang2023OE_31_15697,Xiong2025PRR_7_013238}, phenomenological treatments~\cite{Munoz2014NP_8_550-555,Strekalov2014NP_8_500-501,Chang2016PRL_117_203602a,Munoz2018OO_5_14-26,Bin2018PRA_98_043858a,Bin2020PRL_124_053601,Jiang2023OE_31_15697,Xiong2025PRR_7_013238}, or strict frequency requirements~\cite{Munoz2018OO_5_14-26,Bin2020PRL_124_053601,Bin2021PRL_127_073602,Jiang2023OE_31_15697,Xiong2025PRR_7_013238}, thereby limiting their general applicability and motivating the search for alternative strategies to control photon-bundle statistics. 

Building upon these advances, we investigate the control of photon-bundle statistics within an open quantum systems framework~\cite{Weiss2021__}, where the reduced dynamics is described using a dressed master equation (DME)~\cite{Beaudoin2011PRA_84_043832,settineri2018pra,LeBoite2020AQT_3_1900140a}. 
The traditional inclusion of external driving field is eliminated in this study.
Within this approach, QRM and its extended variants allow efficient manipulation of photon-bundle statistics by tuning only the light-matter coupling strength, in absence of external field modulation. This minimal-control scheme is particularly attractive for realistic finite-temperature environments and provides a viable route toward realizing on-demand photon-bundle sources.
The remainder of this paper is organized as follows. Sec.~\ref{sec:Model and method} presents the QRM framework and its variants, along with the DME approach and the definition of photon-bundle correlation functions. In Sec.~\ref{sec:Photon Bundles Correlation Properties}, we present a systematic analysis of second-order photon-bundle correlations, focusing on the roles of temperature and light-matter coupling strength. We then examine the effects of anisotropic and Stark couplings, as well as their combined influence on photon-bundle statistics and QPTs. In Sec.~\ref{sec:Experimental Feasibility}, we analyze the experimental feasibility of implementing photon-bundle control based on the Rabi family models. Finally, Sec.~\ref{sec: Conclusion} summarizes the main results and discusses potential directions for future research.

\section{Model and Method}
\label{sec:Model and method}
\subsection{Quantum Rabi Model Family}

The QRM describes the fundamental interaction between a two-level system and a single-mode bosonic field~\cite{Rabi1936PR_49_324-328,Braak2016JPAMT_49_300301}. By introducing nonlinear Stark couplings and anisotropic interactions, representative form of the QRM family, encompassing a wider range of light-matter interaction regimes, which can be written in the following 
\begin{equation}
\begin{aligned}
H_{\textnormal{QRMF}} =& \left(\frac{1}{2}\Delta + U a^{\dagger} a\right)\sigma_z + \omega_0 a^{\dagger} a \\&+ g \big[ (a\sigma_+ + a^{\dagger}\sigma_-) + r(a\sigma_- + a^{\dagger}\sigma_+) \big],
\label{H_QRMF}
\end{aligned}
\end{equation}
where $\Delta$ denotes the qubit energy splitting, $U$ is the nonlinear Stark coupling strength, and $a^\dagger$ ($a$) is the photon creation (annihilation) operator of the cavity field with frequency $\omega_0$. The operators $\sigma_\pm = \frac{1}{2}(\sigma_x \pm i \sigma_y)$ represent the excitation and relaxation of qubit between the ground state $|0\rangle$ and the excited state $|1\rangle$, with $\sigma_{\alpha=x,y,z}$ being the Pauli matrices. The parameter $g$ characterizes the qubit-cavity coupling strength, while $r$ controls the contribution of the CRW terms. With experimental access to strong coupling (SC)~\cite{Clarke2008N_453_1031-1042,Askenazi2014NJP_16_043029}, ultrastrong coupling (USC)~\cite{Forn-Diaz2010PRL_105_237001,forn-diazUltrastrongCouplingSingle2017}, and deep strong coupling (DSC) regimes~\cite{Casanova2010PRL_105_263603,Yoshihara2017NP_13_44-47}, the influence of the CRW terms becomes progressively more significant.

This formulation naturally connects AQRSM~\cite{xieFirstorderContinuousQuantum2020} to several well-known models with the appropriate parameter choices. Specifically, setting the Stark coupling to zero ($U=0$) reduces the Hamiltonian to the AQRM~\cite{Xie2014PRX_4_021046,Liu2017PRL_119_,Chen2021PRA_103_043708}. In addition, imposing $r=1$ recovers the standard QRM~\cite{Rabi1936PR_49_324-328,Braak2016JPAMT_49_300301}, in which the CRW terms contribute fully. In the weak-coupling limit, the CRW terms $(a\sigma_- + a^{\dagger}\sigma_+)$ become negligible, leading to the familiar Jaynes-Cummings model (JCM) with $r=0$. These connections highlight how the model in Eq.~\eqref{H_QRMF} provides a unified framework for exploring a range of light-matter interaction regimes.

The system exhibits $Z_2$ symmetry, as indicated by the commutation relation $[\mathit{\Pi}, H_{\textnormal{QRMF}}] = 0$. The conserved parity operator $\mathit{\Pi}$ is defined as $\exp(i\pi N)$, where $N = a^{\dagger}a + \frac{1}{2}(\sigma_z + 1)$ denotes the total excitation number. Its eigenvalues are $\pm 1$, depending on whether $N$ is even or odd. Although the CRW terms break the conservation of the total excitation number, the parity symmetry remains intact, which not only simplifies theoretical treatments but also plays a key role in governing nonclassical phenomena. All numerical results reported in this work were obtained using a sufficiently converged truncation of the photon Fock space at $N_{\textnormal{tr}} = 100$, a setting that will be implicitly assumed throughout the subsequent discussions.

\subsection{Quantum Dressed Master Equation}
In realistic physical systems, quantum coherence is inevitably influenced by the environment~\cite{Weiss2021__}. 
To incorporate dissipation in the QRM family, we assume that the qubit and the cavity field are independently coupled to their respective thermal baths. 
The total Hamiltonian of the composite system is written as
\begin{equation}
H_{\textnormal{total}}=H_{\textnormal{QRMF}}+H_{\textnormal{B}}+V.
\label{H_total}
\end{equation}
The first term is the system Hamiltonian defined in Eq.~\eqref{H_QRMF}. 
The thermal baths are described by
$H_{\textnormal{B}}=\sum_{u=\textnormal{q,c};k}\omega_{k}b_{u,k}^{\dag }b_{u,k}$,
where $b_{u,k}^{\dag }$ ($b_{u,k}$) creates (annihilates) a bosonic excitation with frequency $\omega_k$ in the $u$th bath.
The system-bath interaction is given by $V=V_{\textnormal{q}}+V_{\textnormal{c}}$, corresponding to the qubit and cavity couplings, respectively. 
These interactions take the form
\begin{subequations}
\begin{align}
V_{\textnormal{q}}=&\sum\limits_{k}\lambda _{\textnormal{q},k}(b_{\textnormal{q},k}+b_{\textnormal{q},k}^{\dag })\sigma_{x},\\
V_{\textnormal{c}}=&\sum\limits_{k}\lambda _{\textnormal{c},k}(b_{\textnormal{c},k}+b_{\textnormal{c},k}^{\dag })(a+a^{\dag}),
\label{Interaction}
\end{align}
\end{subequations}
where $\lambda _{\textnormal{q}(\textnormal{c}),k}$ denotes the coupling strength between the qubit (cavity) and its corresponding bath. Each subsystem is characterized by an independent spectral density, which is defined as
$J_{\textnormal{q}(\textnormal{c})}(\omega)=2\pi\sum_{k}|\lambda_{\textnormal{q}(\textnormal{c}),k}|^{2}\delta(\omega-\omega_{k})$.
In this work, we adopt the Ohmic-type spectral function, i.e.,
$J_{\textnormal{q}}(\omega)=\alpha_{\textnormal{q}}\frac{\omega}{\Delta}\exp(-|\omega|/\omega_{c})$
and
$J_{\textnormal{c}}(\omega)=\alpha_{\textnormal{c}}\frac{\omega}{\omega_{0}}\exp(-|\omega|/\omega_{c})$,
where $\alpha_{\textnormal{q}(\textnormal{c})}$ is the dimensionless coupling strength and $\omega_{c}$ is the cutoff frequency.

Under the weak system--bath coupling assumption, we apply the Born--Markov approximation to derive the quantum DME~\cite{Beaudoin2011PRA_84_043832,settineri2018pra,LeBoite2020AQT_3_1900140a}. This approach remains applicable in the USC and DSC regimes, where the qubit and cavity must be treated as a composite system. Consequently, the system--bath interaction is expressed in the eigenbasis $\{|\varphi_n\rangle\}$ of the QRM family as
\begin{subequations}
\begin{align}
V_{\textnormal{q}}=&\sum_{k,m,n}\lambda_{\textnormal{q},k}(b_{\textnormal{q},k}+b_{\textnormal{q},k}^{\dag})P_{nm}^{\textnormal{q}},\\
V_{\textnormal{c}}=&\sum_{k,m,n}\lambda_{\textnormal{c},k}(b_{\textnormal{c},k}+b_{\textnormal{c},k}^{\dag})P_{nm}^{\textnormal{c}},
\end{align}
\end{subequations}
where
$P_{nm}^{\textnormal{q}}=\langle\varphi_n|\sigma_x|\varphi_m\rangle|\varphi_n\rangle\langle\varphi_m|$
and
$P_{nm}^{\textnormal{c}}=\langle\varphi_n|(a^{\dag}+a)|\varphi_m\rangle|\varphi_n\rangle\langle\varphi_m|$.

The resulting quantum DME is given by
\begin{equation}
\begin{aligned}
\frac{\partial \rho}{\partial t}
=&-i[H_{\textnormal{QRMF}},\rho]\\
&+\sum\limits_{j,k>j}^{u=\textnormal{q,c}}
\Big\{
\Gamma_{u}^{j,k}[1+n_{u}(\Delta_{k,j})]\mathcal{D}[|\varphi_j\rangle\langle\varphi_k|,\rho]\\
&+\Gamma_{u}^{j,k}n_{u}(\Delta_{k,j})\mathcal{D}[|\varphi_k\rangle\langle\varphi_j|,\rho]
\Big\}.
\end{aligned}
\label{DME}
\end{equation}
Here,
$\mathcal{D}[\mathcal{O},\rho]=(2\mathcal{O}\rho\mathcal{O}^{\dag}-\mathcal{O}^{\dag}\mathcal{O}\rho-\rho\mathcal{O}^{\dag}\mathcal{O})/2$
is the dissipator and $\rho$ is the reduced density matrix.
The thermal occupation number follows the Bose--Einstein distribution
$n_u(\Delta_{k,j})=[\exp(\Delta_{k,j}/k_{\textnormal{B}}T_u)-1]^{-1}$,
where $\Delta_{k,j}=E_k-E_j$. The transition rates induced by the qubit and cavity baths are
\begin{subequations}
\begin{align}
\Gamma_{\textnormal{q}}^{k,j}
=&\alpha_{\textnormal{q}}\frac{\Delta_{k,j}}{\Delta}
\left|\langle\varphi_j|(\sigma_-+\sigma_+)|\varphi_k\rangle\right|^2 e^{-|\Delta_{k,j}|/\omega_c},\\
\Gamma_{\textnormal{c}}^{k,j}
=&\alpha_{\textnormal{c}}\frac{\Delta_{k,j}}{\omega_0}
\left|\langle\varphi_j|(a+a^{\dag})|\varphi_k\rangle\right|^2 e^{-|\Delta_{k,j}|/\omega_c}.
\end{align}
\end{subequations}
Parity conservation imposes selection rules such that transitions occur only between states of opposite parity.

Using Eq.~\eqref{DME}, the evolution of the density matrix elements $\rho_{m,n}$ is obtained as
\begin{equation}
\begin{aligned}
\dot{\rho}_{m,n}
&= -i\Delta_{m,n}\rho_{m,n}
 + \delta_{m,n}\Bigl[
\sum_{k>m}^{u=\mathrm{q,c}}\!\Gamma_u^{m,k}\bigl(1+n_u(\Delta_{k,m})\bigr)
\\
&\qquad\quad
+ \sum_{k<m}^{u=\mathrm{q,c}}\!\Gamma_u^{k,m}n_u(\Delta_{m,k})
\Bigr]\rho_{k,k} \\
&- \frac{1}{2}
\Bigl[
\sum_{k<m}^{u=\mathrm{q,c}}\!\Gamma_u^{k,m}\bigl(1+n_u(\Delta_{m,k})\bigr)+ \sum_{k>m}^{u=\mathrm{q,c}}\!\Gamma_u^{m,k}n_u(\Delta_{k,m})
\\
&+ \sum_{k<n}^{u=\mathrm{q,c}}\!\Gamma_u^{k,n}\bigl(1+n_u(\Delta_{n,k})\bigr)
+ \sum_{k>n}^{u=\mathrm{q,c}}\!\Gamma_u^{n,k}n_u(\Delta_{k,n})
\Bigr]\rho_{m,n}.
\end{aligned}
\end{equation}
For $m\neq n$, the steady-state solution satisfies $\rho_{m,n}=0$.
For $m=n$, using $1+n_u(\Delta_{k,m})=-n_u(\Delta_{m,k})$ and $\Gamma_u^{m,k}=-\Gamma_u^{k,m}$,
the steady-state condition reduces to
$
\sum_{k\neq m}^{u=\textnormal{q,c}}
\Gamma_u^{k,m}n_u(\Delta_{m,k})\rho_{k,k}
-
\sum_{k\neq m}^{u=\textnormal{q,c}}
\Gamma_u^{k,m}[1+n_u(\Delta_{m,k})]\rho_{m,m}
=0$.
When $T_{\textnormal{q}}=T_{\textnormal{c}}=T$, the system relaxes to thermal equilibrium.
The steady-state density matrix takes the canonical form
$\rho=\sum_n P_n|\varphi_n\rangle\langle\varphi_n|$,
with
$P_n=\mathrm{e}^{-E_n/k_{\textnormal{B}}T}/Z$
and
$Z=\mathrm{Tr}(\mathrm{e}^{-H_{\textnormal{QRMF}}/k_{\textnormal{B}}T})$.
In the zero-temperature limit $T=0$, the steady state collapses to the ground state of the system.

\subsection{Photon Bundle Correlation Functions in the Dressed Picture}
\label{sec:statistics}
Coherence is a fundamental concept in optics, and in the quantum regime, it admits a clear operational interpretation, as demonstrated in~\cite{Glauber1963PR_130_2529-2539,Scully1997__}. When the electromagnetic field is quantized, coherence is naturally characterized by quantum correlation functions of arbitrary order, which provide a powerful tool for identifying nonclassical light fields. Experimentally, these correlation functions are accessed through measurements of the emitted radiation field. In the weak qubit-cavity coupling regime, the output field is approximately proportional to the intra-cavity field via standard input-output theory, such that the conventional definitions of higher-order correlation functions remain applicable.

In the USC regime, this approximation breaks down, as it predicts an unphysical finite photon emission from the ground state.
To resolve this issue, a dressed-state input--output formalism was introduced~\cite{Ridolfo2012PRL_109_193602,Ridolfo2013PRL_110_163601,Ridolfo2013PRA_88_063812,Garziano2013PRA_88_063829}, in which the output field operator is given by
$A_{\mathrm{out}}=A_{\mathrm{in}}-i\sqrt{\kappa}\,X^{+}$.
Here, $\kappa$ denotes the rate of photon leakage in external detection modes, and $X^{+}$ is the
dressed-state detection operator defined by $X^{+}=-i\sum_{j,k>j}\Delta
_{k,j}X_{j,k}\left\vert \varphi _{j}\right\rangle \left\langle \varphi
_{k}\right\vert $. Here, $X_{j,k}=\left\langle \varphi _{j}\right\vert
(a+a^{\dag })\left\vert \varphi _{k}\right\rangle$ are the matrix elements of the detection operator.
For a vacuum input field, the output photon flux is expressed as
$I_{\mathrm{out}}=\kappa\left\langle X^{-}(t)X^{+}(t)\right\rangle$, 
with $X^{-}=(X^{+})^{\dag}$. This formulation eliminates the unphysical ground-state photon current, since $X^{+}|\varphi_{0}\rangle=0$.
Moreover, the matrix elements of $X^{+}$ vanish between eigenstates of the same parity, thereby imposing a parity-based selection rule on radiative transitions.

The $n$th-order photon correlation function in this framework becomes~\cite{Glauber1963PR_130_2529-2539,Scully1997__,Ridolfo2012PRL_109_193602,Ridolfo2013PRL_110_163601}
\begin{equation}
\begin{aligned}
G^{(n)}(\boldsymbol{\tau})
&=
\frac{
\left\langle
\prod_{j=0}^{n-1} X^{-}(t+\tau_j)\,
\prod_{j=n-1}^{0} X^{+}(t+\tau_j)
\right\rangle
}{
\prod_{j=0}^{n-1}
\left\langle (X^{-}X^{+})(t+\tau_j) \right\rangle
}
\end{aligned}
\end{equation}
where $\tau_0\equiv0$, and the products of the operators $X^{-}$ are ordered in time,
with the operators $X^{+}$ written in reverse temporal order. Its steady-state
zero-time-delay form is
\begin{equation}
G^{(n)}(0)=
\frac{
\left\langle (X^{-})^n(X^{+})^n \right\rangle_{\!ss}
}{
\left\langle X^{-}X^{+} \right\rangle_{\!ss}^{\,n}
}.
\end{equation}

Although $G^{(n)}(0)$ already captures essential information about instantaneous photon correlations and provides a standard characterization for isolated
single-photon emission, it does not uniquely determine the correlation properties
of multiphoton bundles. In multiphoton emitters, the elementary emission unit is
no longer a single photon but a bundle consisting of $m$ photons, which
calls for a dedicated correlation function tailored to such composite emission
processes. Consequently, the $m$-photon bundle correlation function is defined
as~\cite{Munoz2014NP_8_550-555,Strekalov2014NP_8_500-501}
\begin{equation}
\begin{aligned}
G_m^{(n)}(\tau)
&=
\frac{
\left\langle
\prod_{j=0}^{n-1}
[X^{-}(t+\tau_j)]^m\,
[X^{+}(t+\tau_{n-1-j})]^m
\right\rangle
}{
\prod_{j=0}^{n-1}
\left\langle
[X^{-}(t+\tau_j)X^{+}(t+\tau_j)]^m
\right\rangle
}.
\end{aligned}
\label{Gmn}
\end{equation}
where the products of the operators $X^{-}$ are ordered in time,
and the operators $X^{+}$ are written in reverse temporal order. This quantity characterizes the time-delayed $n$th-order correlations of $m$-photon bundles. In the steady state, the zero-time-delay limit reduces to
\begin{equation}
G_{m}^{(n)}(0)
=
\frac{
\left\langle (X^{-})^{mn}(X^{+})^{mn} \right\rangle_{\!ss}
}{
\left\langle (X^{-})^{m}(X^{+})^{m} \right\rangle_{\!ss}^{\,n}
},
\label{Gmn0}
\end{equation}
which serves as a natural generalization of the conventional photon
correlation function to the multiphoton-bundle regime.

Although the above definitions apply to arbitrary order $n$, the second-order correlation already provides the simplest and most direct characterization of $m$-photon bundles. The general expression of its time-delay correlation function reduces to
\begin{equation}
G_m^{(2)}(\tau)
=
\frac{
\left\langle
[X^{-}(t)]^m
[X^{-}(t+\tau)]^m
[X^{+}(t+\tau)]^m
[X^{+}(t)]^m
\right\rangle
}{
\left\langle
[X^{-}(t)X^{+}(t)]^m
\right\rangle^{2}
},
\label{tau}
\end{equation}
where $\langle O(t) \rangle = \mathrm{Tr}\{O\,\rho(t)\}$ denotes the expectation value. The density operator obeys the master equation
$\frac{d}{dt}\rho(t) = \mathcal{L}\rho(t)$,
where the Liouvillian superoperator $\mathcal{L}$ is specified in Eq.~\eqref{DME}.
The multi-time correlation functions are evaluated according to the quantum regression theorem~\cite{carmichaelStatisticalMethodsQuantum2008},
\begin{equation}
\langle O_1(t) O_2(t+\tau) O_3(t) \rangle
=
\mathrm{Tr}
\left\{
 O_2 e^{\mathcal{L}\tau}
\left[ O_3 \rho(t) O_1\right]
\right\}.
\end{equation}

In this case, the emission of $m$-photon bundles is termed antibunched if $G^{(2)}_m(0) < G^{(2)}_m(\tau)$, and bunched if $G^{(2)}_m(0) > G^{(2)}_m(\tau)$, within the well-defined region $\tau > \tau_{\mathrm{min}}$. In the long-time limit,
$G^{(2)}_m(\tau)\to 1$. Accordingly, $G^{(2)}_m(0)<1$ indicates antibunching of $m$-photon bundles and a tendency toward sequential emission, while $G^{(2)}_m(0)>1$ signals an enhanced probability for bundle emission to occur close in time.

For the special case $m=1$, $G^{(2)}_1(0)$ reduces to the standard second-order correlation function $G^{(2)}(0)$, which fully characterizes the zero-delay statistics of single-photon emission. However, for $m>1$, the zero-time delay $G^{(2)}_1(0)$ alone is generally insufficient to uniquely characterize the underlying multiphoton emission processes. In this situation, a combined analysis of $G^{(2)}_1(0)$, $G^{(2)}_2(0)$ provides a more complete classification of photon-emission regimes. In particular, different combinations of $G^{(2)}_1(0)$ and $G^{(2)}_2(0)$ correspond to different statistical tendencies, which can be summarized as follows:
\begin{itemize}

\item $G_1^{(2)}(0) > 1$ and $G_2^{(2)}(0) > 1$: 
Both single photon and two-photon bundles exhibit bunching. The overall photon emission resembles a laser-type continuous output, where multiple minimal emission units coexist within temporally overlapping clusters.

\item $G_1^{(2)}(0) > 1$ and $G_2^{(2)}(0) < 1$: 
Single photons are bunching, while two-photon bundles exhibit antibunching at zero time delay, indicating temporally separated bundle emission. This behavior corresponds to a gun-type sequential two-photon source.

\item $G_1^{(2)}(0) < 1$ and $G_2^{(2)}(0) < 1$: 
Both single photon and two-photon bundles are antibunched,
with emissions occurring in a discrete and spaced manner,
indicating a strong photon blockade effect beyond the conventional single photon blockade.

\item $G_1^{(2)}(0) < 1$ and $G_2^{(2)}(0) > 1$: Single photons exhibit antibunching, while two-photon bundles are emitted consecutively. This phenomenon seems counterintuitive, but it originates from path competition and interference effects: single-photon antibunching arises from destructive interference between competing paths, representing a nontraditional photon blockade~\cite{Flayac2017PRA_96_053810}, while photon-bundle bunching is stabilized through consecutive transition channels. In this case, the system behaves as a genuine two-photon bundle laser, with each photon pair serving as a minimal emission bundle.

\end{itemize}

Although bundles with larger emission units can, in principle, give rise to diverse emission behavior, they are generally closely linked to the single- and two-photon correlation functions. Therefore, in this work, we focus on two-photon bundles. Among the above regimes, our
primary interest lies in enhancing photon bundle emission and controlling two-photon bundles to access both gun-type sequential and laser-type continuous emission behaviors.

\section{Photon Bundles Correlation Properties for the Quantum Rabi Model Family}
In this chapter, we investigate two-photon bundle correlations at finite temperatures, based on steady-state solutions obtained from DME at Eq.~\eqref{DME}. A systematic analysis is carried out to examine the effects of temperature and coupling strength on photon bundle correlation behaviors. Starting from QRM, we incorporate anisotropy and Stark effects to explore their influence on two-photon bundle correlations. By comprehensively characterizing the photon-bundle statistical properties of the QRM family, we elucidate how intrinsic system parameters govern the correlation behaviors, thereby providing a direct and effective approach for controlling photon-bundle correlations~\cite{Munoz2014NP_8_550-555,Strekalov2014NP_8_500-501}.

\label{sec:Photon Bundles Correlation Properties}
\subsection{Quantum Rabi Model}
We begin by considering the simplest quantum model describing light-matter interaction, namely QRM~\cite{Rabi1936PR_49_324-328,Braak2016JPAMT_49_300301}. Starting from Eq.~\eqref{H_QRMF}, by setting $r = 1$ and $U = 0$, the Hamiltonian of the standard QRM is obtained as
\begin{equation}
H_{\textnormal{QRM}} = \frac{\Delta}{2}\sigma_z + \omega_0 a^{\dagger} a + g \sigma_x\left(a + a^{\dagger}\right),
\label{H_QRM}
\end{equation}
where the parameters are defined as in Eq.~\eqref{H_QRMF}. Previous studies have demonstrated that photon bundles can be engineered through various mechanisms, including external laser driving, multiphoton coupling schemes, enhanced dissipation strength, or tailored dissipation channels~\cite{Munoz2014NP_8_550-555,Strekalov2014NP_8_500-501,Chang2016PRL_117_203602a,Munoz2018OO_5_14-26,Bin2018PRA_98_043858a,Bin2020PRL_124_053601,Bin2021PRL_127_073602,Jiang2023OE_31_15697,Xiong2025PRR_7_013238}. Here, we instead consider a physically natural scenario at finite temperature and examine whether photon bundles can be induced solely by tuning the light-matter coupling strength. 

\subsubsection{Zero-Time Delay Correlation Properties of Photon Bundles}
\begin{figure}[htbp]
\centering\includegraphics[width=8.5cm]{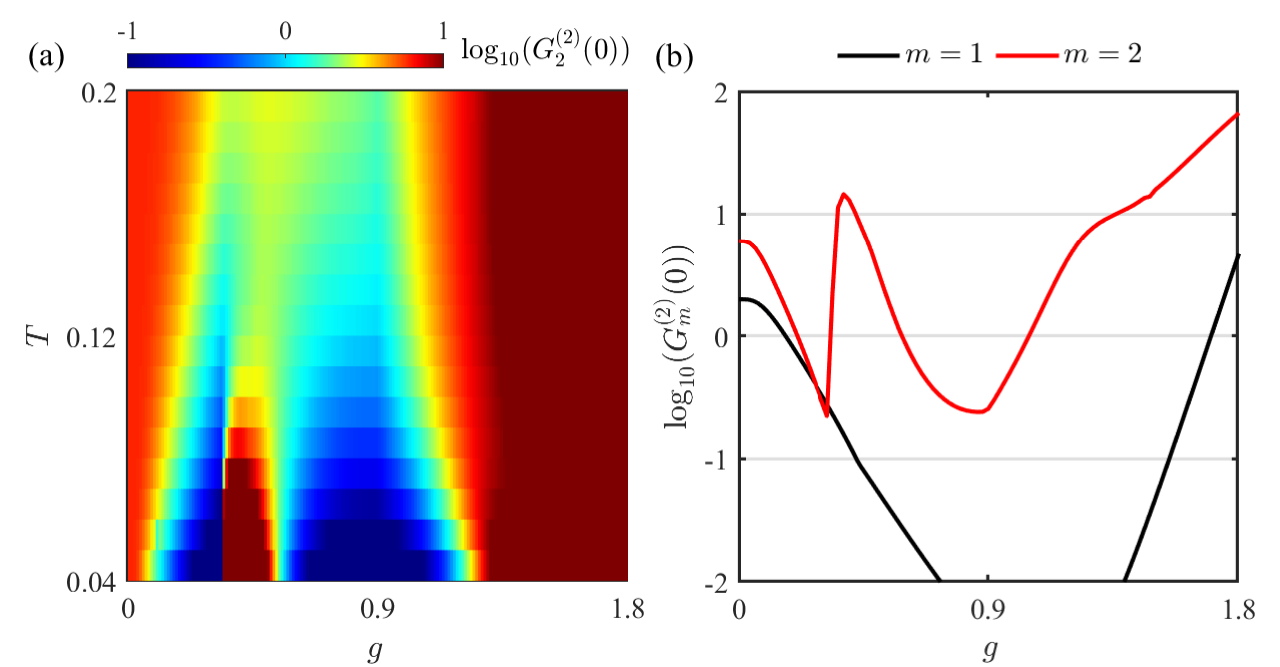}
\caption{Photon correlation properties in the QRM. (a) Logarithm of the zero-delay two-photon bundles correlation $G_2^{(2)}(0)$ versus temperature $T$ and coupling strength $g$. (b) Logarithm of the zero-delay $m$-photon bundles correlation $G_m^{(2)}(0)$ at $T=0.07$ as a function of $g$. Parameters: $\Delta=1$, $\omega_0=1$, $r=1$, $U=0$, $\alpha_{\mathrm{c}}=\alpha_{\mathrm{q}}=10^{-3}$, $\omega_{\mathrm{c}}=10\omega_0$.}
\label{Fig1}
\end{figure}
To characterize how photon bundle correlations can be controlled by temperature and light-matter coupling strength, we first focus on the zero-time delay second-order correlation of two-photon bundles, which characterizes the instantaneous statistics and the dominant emission channels. Figure~\ref{Fig1}(a) illustrates the behavior of the correlation function defined in Eq.~\eqref{Gmn0} as a function of temperature $T$ and coupling strength $g$.

At low temperatures, as the coupling strength $g$ increases, the photon bundle behavior undergoes two successive transitions between bunching and antibunching, following a sequence of bunching $\rightarrow$ antibunching $\rightarrow$ bunching $\rightarrow$ antibunching $\rightarrow$ bunching, thereby revealing multiple distinct parameter regimes associated with different bundle correlation properties. With rising temperature, the photon bundle bunching regimes progressively expand, leading to stronger photon bundle bunching across a broader parameter space. As predicted, the formation of photon bundles can be effectively driven by temperature and coupling strength $g$, highlighting their role in tuning and controlling photon bundle correlations.
At sufficiently high temperature, photon-bundle bunching emerges over nearly the entire parameter regime. 

To better control the photon-bundle correlations, we fixed the temperature at $T = 0.07$ for subsequent analysis. 
Fig.~\ref{Fig1}(b) displays the zero-time delay second-order correlation functions for single-photon and two-photon bundle processes in QRM, corresponding to $m=1$ and $m=2$, respectively. For brevity, we refer to these zero-delay second-order correlations as the single-photon and two-photon bundle correlations.

We first examine the single-photon correlation function. The photon statistics evolve continuously through a successive transition, namely bunching $\rightarrow$ antibunching $\rightarrow$ bunching. Specifically, single photon bunching, characterized by $G_1^{(2)}>1$, appears only near the two ends of the coupling range, where the coupling strength takes relatively small or relatively large values. In contrast, over most of the intermediate parameter region, the system exhibits single-photon antibunching, where photons are emitted individually with clear temporal separation.

Turning to the two-photon bundle correlation function, a richer and more intricate statistical evolution is observed. The two-photon bundle experiences two complete transitions, each characterized by a bunching $\rightarrow$ antibunching $\rightarrow$ bunching pattern.
In particular, in parameter regions where single-photon bunching occurs, two-photon bundle simultaneously exhibits bunching behavior ($G_1^{(2)} > 1$ and $G_2^{(2)} > 1$). This suggests that bunched photon emission in these regimes includes both single-photon and two-photon bundles as fundamental emission units, while other types of emission unit may also be present. Moreover, in certain parameter regions originally characterized by single-photon antibunching, the system also displays two-photon bundle antibunching ($G_1^{(2)} < 1$ and $G_2^{(2)} < 1$). Photon emission in these regimes occurs under strong photon blockade, indicating that the emission involves isolated single photons as well as sequentially emitted two-photon bundles as fundamental emission units.

Beyond these regimes, we further identify parameter regions satisfying $G_1^{(2)} < 1$ while $G_2^{(2)} > 1$, photon emission exhibits two distinct channels: single-photon events are antibunched, reflecting photon blockade, whereas two-photon bundles are emitted sequentially, each forming a minimal emission unit of a genuine two-photon bundle laser. This phenomenon may seem counterintuitive at first, as single-photon antibunching is typically thought to suppress higher-order bunching behavior. However, its physical origin can be traced back to competition effects between multi-transition channels: at the single-photon level, quantum interference between different transition paths leads to the cancellation of emission probabilities, resulting in a nontraditional photon blockade mechanism, manifested as single-photon antibunching. In contrast, within the photon bundle bunching channels, this path competition is effectively circumvented, and the two photons form stable correlated emissions through consecutive transition channels, maintaining significant bunching characteristics.

Therefore, within the framework of the QRM, we find that, by solely tuning the light-matter coupling strength, the zero-time-delay correlation statistics of two-photon bundles can be
effectively controlled at finite temperatures. In particular, in certain coupling regimes, pronounced two-photon bundle bunching is observed, which is indicative of emission behavior
relevant to two-photon-bundle lasers, where the fundamental emission units are photon bundles. In contrast, in other parameter regimes, strong two-photon bundle antibunching emerges, suggesting emission characteristics compatible with a bundle-based quantum gun.

\subsubsection{Time-Delay Correlation Properties of Photon Bundles}
\begin{figure}[htbp]
\centering\includegraphics[width=8.5cm]{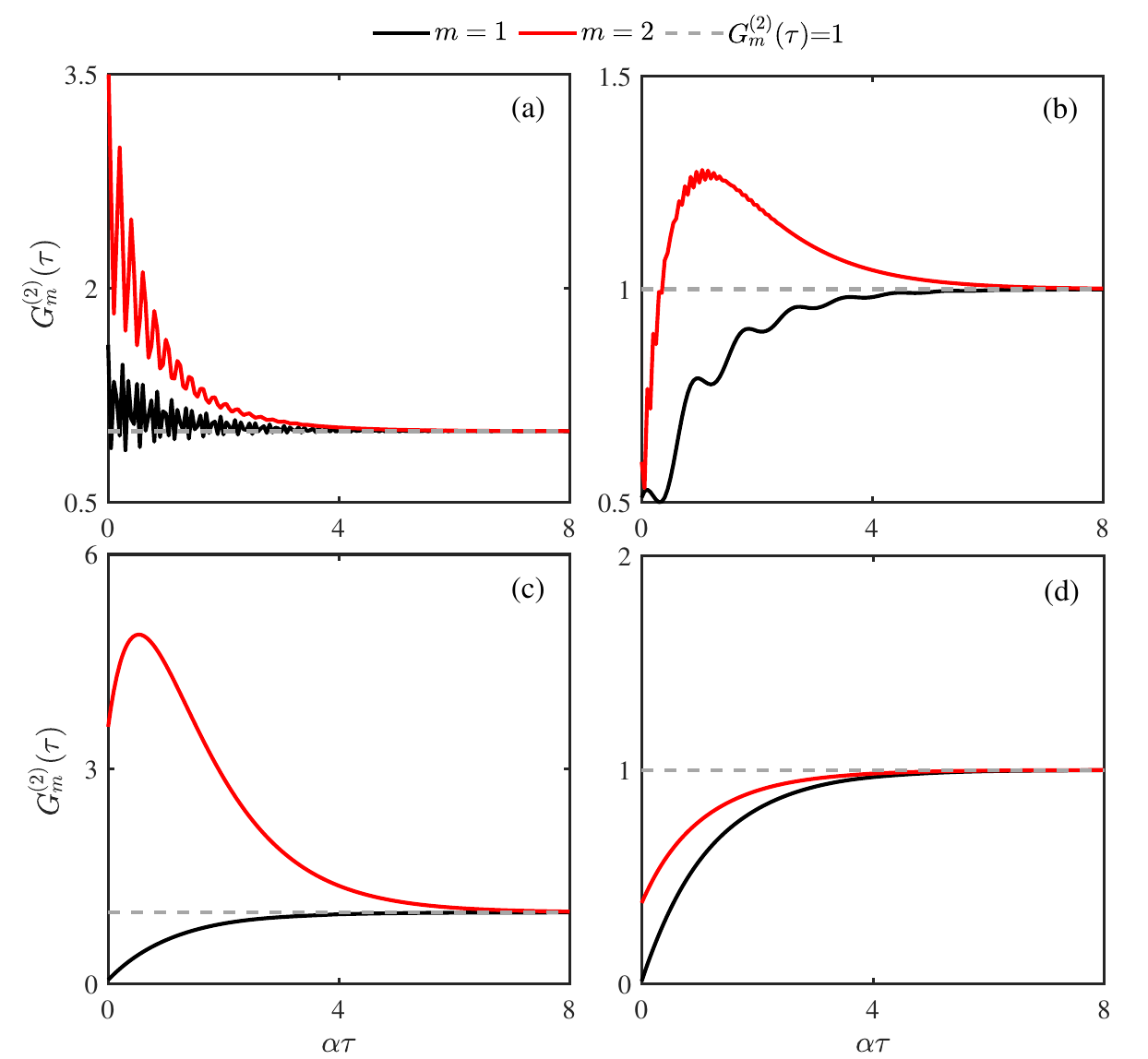}
\caption{The time-delayed $m$-photon bundles correlation $G_m^{(2)}(\tau)$ versus renormalized time  $\alpha\tau$ with $\alpha=\alpha_c$ for the QRM. (a) $g=0.1$; (b) $g=0.25$; (c) $g=0.5$; (d) $g=0.7$. Other parameters are set as Fig.~\ref{Fig1}(b).}
\label{Fig12}
\end{figure}
We employ the time-delayed $m$-photon-bundle correlation function at Eq.~\eqref{tau} to examine the behaviors of photon-bundle correlations with finite time delay under different light-matter coupling strengths, as shown in Fig.~\ref{Fig12}. The finite time-delayed correlation function $G_m^{(2)}(\tau)$ provides access to temporal ordering and dynamical memory between emission processes, thus supplying time-scale information that cannot be uniquely inferred from the zero-delay result alone. Overall, although the detailed dynamical behaviors differ among the subfigures in Fig.~\ref{Fig12}, all correlations ultimately relax to the uncorrelated steady state of photons in the long-delay limit, i.e., $G_m^{(2)}(\tau{\rightarrow}\infty)\to 1$.

Specifically for $g=0.1$ in Fig.~\ref{Fig12}(a),  the temporal correlations exhibit pronounced oscillatory behaviors, reflecting long-lived dynamical memory. The single-photon correlation predominantly satisfies $G_1^{(2)}(\tau)>1$.
Meanwhile, the correlation of the two-photon bundle also remains $G_2^{(2)}(\tau)>1$ throughout the delay range. This behavior indicates that two-photon bundles are preferentially emitted in a correlated manner over finite delay time, although the correlation strength gradually decays and eventually approaches the uncorrelated steady state. 

For the case $g=0.25$ shown in Fig.~\ref{Fig12}(b), the enhanced qubit-photon coupling significantly suppresses the oscillation amplitude of the temporal correlations both for 
single-photon and two-photon bundle correlations, compared to case $g=0.1$. Importantly, time-delayed correlations reveal a qualitative evolution that is not captured by the zero-delay picture alone: as $\tau$ increases, the two-photon bundle correlation evolves from pronounced antibunching to bunching, whereas the single-photon correlation remains antibunched over the broad delay range. This indicates that the single-photon channel persistently exhibits short-time repulsion, while correlations between successive two-photon bundles gradually change from suppression to enhancement, signaling the buildup of temporally ordered correlations between photon bundles. In the long-delay limit, all correlations are expected to relax to the uncorrelated steady state.

When the qubit-photon coupling increases to $g = 0.5$ in Fig.~\ref{Fig12}(c), the oscillatory features in the temporal correlations nearly disappear, and the system exhibits a smoother relaxation profile. In this regime, the two-photon-bundle correlation remains strongly bunched over a wide delay range, while the single-photon correlation remains antibunched. This indicates that the system suppresses single-photon emission but maintains enhanced correlations for photon bundles, favoring temporally ordered photon bundle emission. With a further increase in coupling to $g = 0.7$ in Fig.~\ref{Fig12}(d), the USC regime further modifies the dynamics, and the single-photon and two-photon bundle correlations evolve monotonically from antibunching to the uncorrelated limit as the delay time increases.

The time-delayed correlations reveal how photon emission processes develop temporal ordering and dynamical memory over time before eventually losing correlation. These results demonstrate that by adjusting the light-matter coupling strength, one can control both the temporal structure and sequential emission dynamics of photon bundles, emphasizing the system's high tunability.

\subsection{Anisotropic Quantum Rabi Model}
\begin{figure}[htbp]
\centering\includegraphics[width=8.5cm]{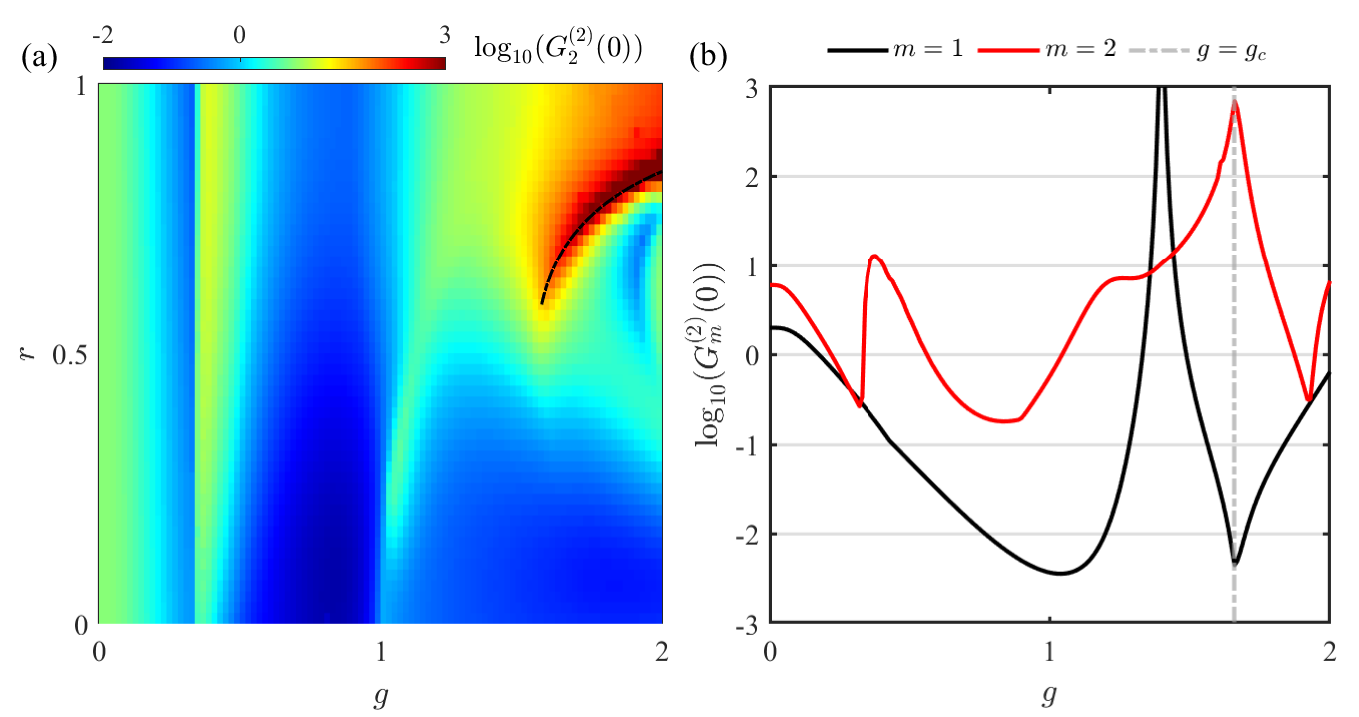}
\caption{Photon correlation properties in the AQRM. (a) Logarithm of the zero-delay two-photon bundles correlation $G_2^{(2)}(0)$ versus anisotropic parameter $r$ and coupling strength $g$. (b) Logarithm of the zero-delay $m$-photon bundles correlation $G_m^{(2)}(0)$ at $r=0.7$ as a function of $g$. Other parameters are set as Fig.~\ref{Fig1}(b).}
\label{Fig2}
\end{figure}
When the RW and CRW terms contribute with distinct strengths, the model in Eq.~\eqref{H_QRMF} evolves into the AQRM~\cite{Xie2014PRX_4_021046,Liu2017PRL_119_,Chen2021PRA_103_043708} under condition $U = 0$. Its Hamiltonian can be written as
\begin{equation}
H_{\textnormal{AQRM}} = \frac{\Delta}{2}\,\sigma_z + \omega_0 a^{\dagger} a + g \big[ (a \sigma_+ + a^{\dagger} \sigma_-) + r (a \sigma_- + a^{\dagger} \sigma_+) \big],
\label{H_AQRM}
\end{equation}
where the parameters are defined as in Eq.~\eqref{H_QRMF}. 
The two-photon bundle correlation functions are shown in Fig.~\ref{Fig2}(a). We find that the introduction of anisotropy dramatically enhances the crossover behavior of the two-photon bundle correlation function. In contrast to the QRM in Fig.~\ref{Fig1}(b), where only two transitions occur, the two-photon bundle statistics exhibit three successive transitions, each following a bunching $\rightarrow$ antibunching $\rightarrow$ bunching sequence, revealing a qualitatively richer correlation structure. Moreover, strong anisotropy extends the parameter regions exhibiting two-photon bundle antibunching, highlighting the enhanced accessibility of nonclassical photon-pair emission. As anisotropy decreases, the pronounced photon-bundle bunching emerges in the DSC regime for $g>1$.

Fig.~\ref{Fig2}(b) shows a cross section of the $m$-photon bundle correlation function in $r=0.7$, where the two-photon bundle correlation statistics display clear bunching–antibunching transitions, becoming markedly richer than those in the isotropic case (Fig.~\ref{Fig1}(b)). More importantly, a striking feature appears near the characteristic coupling point $g = g_c$~\cite{Chen2021PRA_103_043708}: while the single-photon correlation function exhibits pronounced antibunching, the two-photon bundle correlation function shows anomalously strong bunching, manifested as a sharp, nearly divergent peak. This pronounced contrast between single- and two-photon statistics indicates that parameter regimes near the critical coupling point are particularly favorable for realizing a high-purity two-photon bundle laser.

In particular, the coupling strength $g_c$ coincides with the level-crossing point between the second and third excited states, corresponding to a first-order excited state phase transition, which can be identified either through exact numerical diagonalization or by analytic Bogoliubov transformation~\cite{Chen2021PRA_103_043708}. In the strong-bunching region of Fig.~\ref{Fig2}(a), we also find that this region perfectly aligns with the excited-state phase transition interval, indicated by the black dashed lines. This observation implies that the two-photon bundle correlation function provides a more direct and robust diagnostic for excited-state phase transitions than methods that rely solely on single-photon statistics. Specifically, at the first-order transition point, it exhibits anomalous bunching, while the single-photon correlation function shows antibunching—a clear qualitative contrast that transcends mere quantitative variations in single-photon statistics.


\subsection{Quantum Rabi–Stark Model}
\begin{figure}[h]
\centering\includegraphics[width=8.5cm]{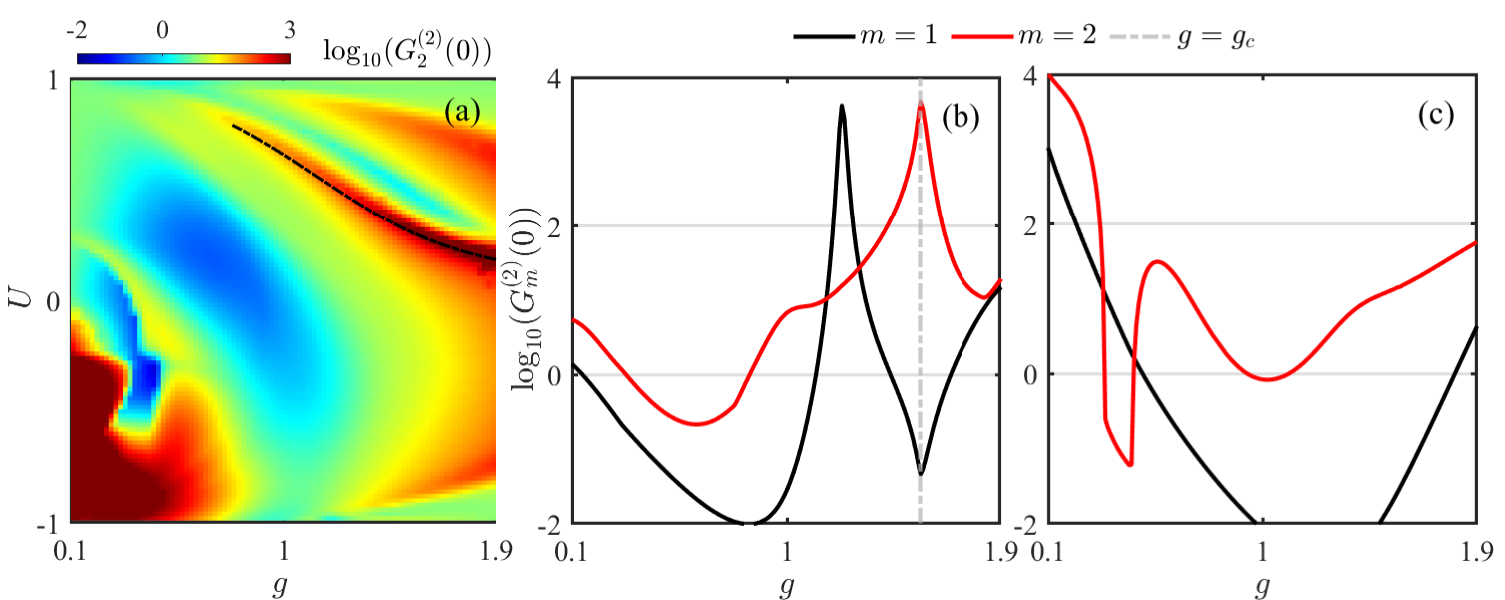}
\caption{Photon correlation properties in the QRSM. (a) Logarithm of the zero-delay two-photon bundles correlation $G_2^{(2)}(0)$ versus Stark parameter $U$ and coupling strength $g$. Logarithm of the zero-delay $m$-photon bundles correlation $G_m^{(2)}(0)$ as a function of $g$ at (b) $U=0.3$ and (c) $U=-0.4$. Other parameters are set as Fig.~\ref{Fig1}(b).}
\label{Fig3}
\end{figure}
Building on our previous findings that anisotropy in the AQRM gives rise to rich two-photon bundle correlation phenomena, we now consider the effects of Stark coupling~\cite{Grimsmo2013PRA_87_033814,Grimsmo2014PRA_89_}. Setting $r = 1$ in the model of Eq.~\eqref{H_QRMF}, the basic form of the QRSM is obtained. Its Hamiltonian can be written as
\begin{equation}
H_{\textnormal{QRSM}} = \left(\frac{1}{2}\Delta + U a^{\dagger} a\right)\sigma_z + \omega_0 a^{\dagger} a + g \sigma_x \left(a + a^{\dagger}\right),
\label{H_QRSM}
\end{equation}
where the parameters are identical to those in Eq.~\eqref{H_QRMF}. 

 It is evident that the introduction of the Stark coupling significantly enriches the two-photon bundle bunching behavior in Fig.~\ref{Fig3}(a). Specifically, for positive Stark coupling strong bunching predominantly occurs in the DSC regime.
Moreover, we note that for positive Stark coupling the regions of strong bunching align precisely with the first-order excited-state phase transition indicated by the black dashed lines. However, for negative Stark coupling, pronounced bunching appears extensively at relatively weaker coupling strengths. This observation further confirms the potential of using anomalous multi-photon bundle correlations as a direct diagnostic for excited-state phase transitions.

Fig.~\ref{Fig3}(b) present the $m$-photon bundle correlation function for positive Stark terms (e.g., $U=0.3$). The introduction of a positive Stark term significantly reduces the richness of photon-bundle statistical transitions by merging two previously isolated antibunching regions into a single continuous one, resulting in only one successive bunching $\rightarrow$ antibunching $\rightarrow$ bunching transition. At the same time, we observe that across a wide region of parameter space, two-photon bundles display markedly stronger bunching compared to the QRM in Fig.~\ref{Fig1}(b), highlighting the distinct role of the positive Stark term in promoting two-photon bundle bunching. In particular, a pronounced correspondence between single photon and photon-bundle correlations persists at the critical coupling strength $g = g_c$~\cite{Xie2019JPAMT_52_245304}, where the single photon correlation function reaches its minimum and the two-photon bundle correlation function simultaneously exhibits a clear maximum. 

In contrast, Fig.~\ref{Fig3}(c) illustrate the case of negative Stark terms ($U=-0.4$).
The two-photon bundle correlation is also dominated by pronounced bunching over almost the entire parameter space. 
In contrast to the $U>0$ case, where the strong two-photon bundle bunching is mainly concentrated in the $g>1$ region, the strong bunching for $U<0$ predominantly appears in the $g<1$ region. More intriguingly, a distinct region characterized by $G_1^{(2)}>1$ and $G_2^{(2)}<1$ emerges, where two-photon bundles are emitted in temporally separated bursts, exhibiting antibunching at the bundle level. Each bundle precisely contains two photons, effectively forming a gun-type sequential two photon source, a phenomenon entirely absent in the AQRM. Remarkably, these behaviors arise solely from finite-temperature effects and adjustable light–matter coupling, without requiring other additional mechanisms.


\subsection{Anisotropic Quantum Rabi–Stark Model}
\begin{figure}[htbp]
\centering\includegraphics[width=8.5cm]{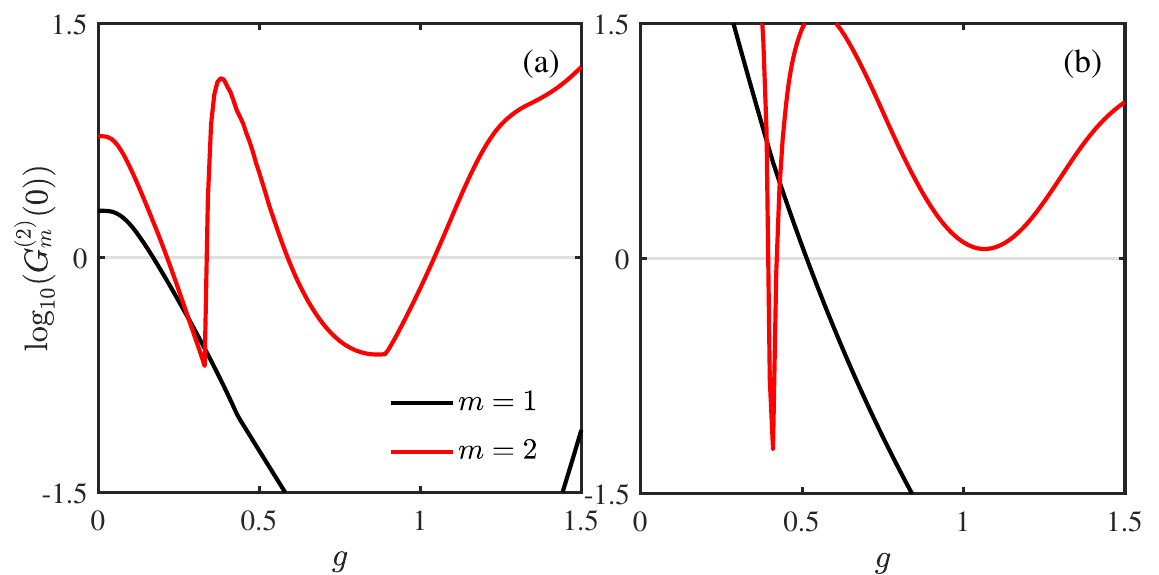}
\caption{Logarithm of the zero-delay $m$-photon bundle correlation function $G_m^{(2)}$ as a function of the coupling strength $g$ for the AQRSM. (a) $U=0$ and $r=0.9$; (b) $U=-0.4$ and $r=0.9$. Other parameters are set as in Fig.~\ref{Fig1}(b).}
\label{Fig4}
\end{figure}
Building on our previous studies, which demonstrated that the independent effects of anisotropy and Stark coupling in the QRM significantly enrich two-photon bundle correlation statistical behaviors, we now investigate their combined effects in AQRSM~\cite{xieFirstorderContinuousQuantum2020}. Taking both anisotropy and Stark coupling into account, the model assumes the same general form as in Eq.~\eqref{H_QRMF}, except that conditions $r \neq 1$ and $U \neq 0$. Specifically, the Hamiltonian can be written as
\begin{equation}
\begin{aligned}
H_{\textnormal{AQRSM}} =& \left(\frac{1}{2}\Delta + U a^{\dagger} a\right)\sigma_z + \omega_0 a^{\dagger} a \\&+ g \big[ (a\sigma_+ + a^{\dagger}\sigma_-) + r (a\sigma_- + a^{\dagger}\sigma_+) \big].
\label{H_AQRSM}
\end{aligned}
\end{equation}

When only anisotropy is included in Fig.~\ref{Fig4}(a), the two-photon bundle correlation statistics only exhibit a single complementary behavior, with $G_1^{(2)}<1$ and $G_2^{(2)}>1$, indicating that photons are emitted either as a single-photon gun or as laser-type emission composed of two-photon bundles. Upon further introducing the Stark coupling in Fig.~\ref{Fig4}(b), previously inaccessible regimes of gun-type two-photon emission emerge, which is characterized by $G_1^{(2)}>1$ and $G_2^{(2)}<1$.
Therefore, it further enhances the controllability and tunability of two-photon bundle emission.

\begin{figure}[htbp]
\centering\includegraphics[width=8.5cm]{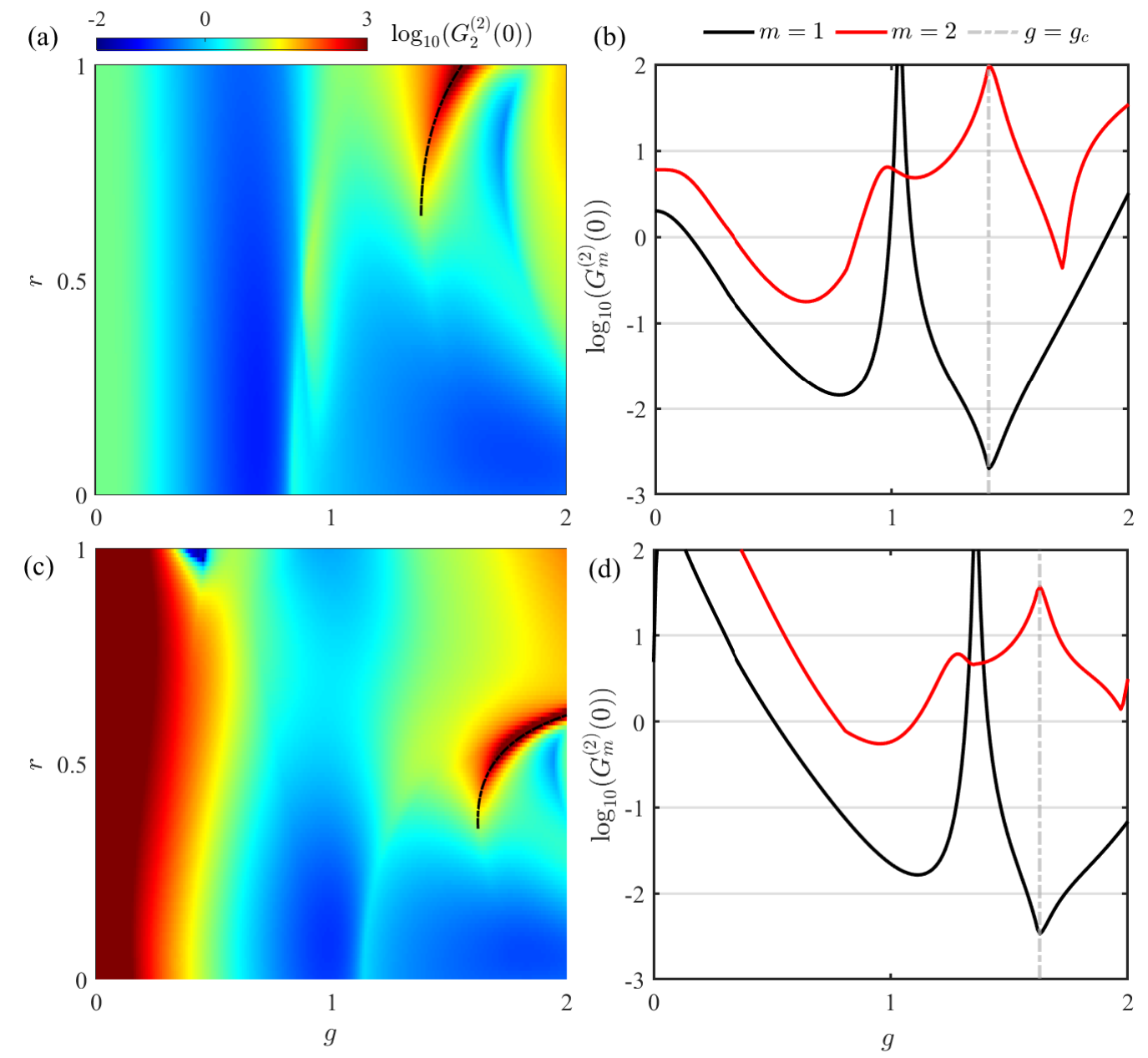}
\caption{Logarithm of the zero-delay two-photon bundles correlation function $G_m^{(2)}$ as a function of anisotropy parameter $r$ and coupling strength $g$. (a) $U=0.3$; (b) $U=0.3$, $r=0.8$; (c) $U=-0.3$; (d) $U=-0.3$, $r=0.4$. Other parameters are set as Fig.~\ref{Fig1}(b).}
\label{Fig5}
\end{figure}

Fig.~\ref{Fig5} exhibits a more comprehensive picture of the parameter space for two-photon bundle correlations. Fig.~\ref{Fig5}(a) correspond to positive Stark couplings, while Fig.~\ref{Fig5}(c) correspond to negative Stark couplings. 
Overall, for $U>0$ the system exhibits rich photon bundle antibunching, suggesting the possibility of realizing gun-type two-photon emission. In contrast, for $U<0$ photon bundle bunching dominates, enabling the implementation of laser-type two-photon bundle emission.

Furthermore, positive Stark terms significantly expand the bundle antibunching regions, especially under strong anisotropy and large coupling strengths. For moderate anisotropy ($0.5 < r < 1$), pronounced photon-bundle bunching appears in the DSC regime. These strong-bunching regions are tied to the level crossing between the second and third excited states, indicating a first-order excited state phase transition (black dashed–dotted line). The representative cut in Fig.~\ref{Fig5}(b) shows a complementary relationship between single-photon antibunching and two-photon bundle bunching, highlighting their joint correlations as a sensitive probe for excited-state phase transitions. Notably, when the system is tuned near the critical coupling $g = g_c$, the coexistence of suppressed single-photon correlations and enhanced two-photon bundle correlations marks a favorable parameter window for stabilizing highly pure two-photon bundle laser emission.

In contrast, for negative Stark terms, two-photon bundle correlations are dominated by strong bundle bunching across most of the parameter space. In the weak-coupling regime ($g<0.5$), bundle bunching is especially prominent, while in the DSC regime ($1<g<2$), this bunching remains strong and follows the excited-state phase transition regions, as shown by the black dashed–dotted line in Fig.~\ref{Fig5}(c). A representative parameter cut in Fig.~\ref{Fig5}(d) shows that single-photon and two-photon bundle correlations respond differently to system parameters, revealing distinct features of the excited-state phase transition. Near the critical coupling point $g=g_c$, the strong two-photon bundle correlations highlight a regime favorable to high-purity two-photon bundle lasing. 

When both anisotropy and Stark coupling are introduced simultaneously, the QRM, which previously exhibited only two successive bunching $\rightarrow$ antibunching $\rightarrow$ bunching transitions, can now undergo three or even more successive transitions. By modulating the anisotropy parameter and Stark coupling appropriately, we can achieve effective control over photon emission, which allows for gun-type and laser-type two-photon bundle emission. This tunability arises from the interplay between anisotropy and Stark-induced level shifts, which reshapes the photon correlation landscape and creates previously inaccessible emission regimes. As a result, one can selectively generate temporally separated photon bundles or clustered laser-type output.


\section{Experimental Feasibility}
\label{sec:Experimental Feasibility}
In the preceding sections, we investigated the controllable generation and manipulation of two-photon bundles in a family of quantum Rabi models. By tuning the light–matter coupling strength, anisotropy, and Stark interaction, we showed that photon-bundle correlations can be systematically engineered, enabling controllable transitions between bunching and antibunching regimes across different parameter ranges. 
From an experimental perspective, the models considered here represent complementary routes for engineering multiphoton emission in strongly-coupled light-matter systems. In the following, we summarize the experimental feasibility of these control mechanisms, based on existing theoretical and experimental studies.

For the QRM, tunable light--matter coupling strengths have been realized across a variety of experimental platforms, spanning from the conventional SC, USC to DSC limits. In the optical domain, SC has been observed in quantum dot-cavity semiconductor architectures~\cite{englundControllingCavityReflectivity2007,hennessyQuantumNatureStrongly2007}, with coupling strengths $g/2\pi \sim 10$–$20\,\mathrm{GHz}$. 
For typical optical frequencies $\omega/2\pi \sim 300\,\mathrm{THz}$, this corresponds to $g/\omega \sim 10^{-4}$, placing these platforms well within the conventional SC regime. By contrast, terahertz intersubband polariton systems in quantum well-plasmonic resonator structures have achieved $g/\omega \sim 0.2$–$0.3$~\cite{Askenazi2014NJP_16_043029}. 
In superconducting circuit QED, a landmark experiment by Niemczyk \textit{et al.} demonstrated normalized coupling strengths as high as $g/\omega \approx 0.12$, clearly entering the USC regime~\cite{niemczykCircuitQuantumElectrodynamics2010}. 
Further advances in circuit engineering ultimately enabled access to the DSC regime, as demonstrated by Yoshihara \textit{et al.}, who achieved $g/\omega$ ratios between $0.72$ and $1.34$ in a flux-qubit-oscillator circuit~\cite{Yoshihara2017NP_13_44-47}.
A comprehensive overview of USC and DSC physics is provided in Ref.~\cite{forn-diazUltrastrongCouplingRegimes2019}. Trapped-ion systems offer another highly controllable route to QRM realizations, where laser-driven interactions generate effective Hamiltonians with widely tunable coupling strengths, as reviewed in Ref.~\cite{leibfriedQuantumDynamicsSingle2003}. In particular, trapped-ion quantum simulators have enabled access to the DSC regime with effective coupling ratios $g/\omega > 1$~\cite{lvQuantumSimulationQuantum2018}. Taken together, these advances establish that the light-matter coupling strength in QRM implementations can be continuously tuned across multiple coupling regimes. 
This broad experimental accessibility provides a solid foundation for exploring QRM-related nonclassical photon statistics and other genuine quantum-optical phenomena beyond the rotating-wave approximation.

Existing studies indicate that anisotropic light-matter interactions beyond the symmetric quantum Rabi limit can be engineered on cavity and circuit QED platforms. 
The anisotropic quantum Rabi model and its tunable coupling structure have been systematically analyzed~\cite{Xie2014PRX_4_021046}. 
Unequal weights of rotating and counter-rotating interaction channels can, in principle, be accessed through periodic frequency-modulation techniques in circuit QED settings~\cite{wangSimulatingAnisotropicQuantum2019a}, 
where the anisotropy ratio $r$ can be continuously tuned by varying the driving amplitude, enabling access to the JC limit ($r\to 0$), the symmetric Rabi point ($r\approx 1$), and the anti-JC regime ($r\to\infty$). 
Circuit-engineering approaches enabling independent coupling tuning to different field quadratures have also been proposed in superconducting architectures~\cite{baksicControllingDiscreteContinuous2014}. Moreover, Rabi-type qubit-oscillator effective descriptions can also arise in spin--orbit coupled semiconductor structures, where the relative strengths of Rashba and Dresselhaus couplings provide a natural source of anisotropy and tunability~\cite{erlingssonEnergySpectraQuantum2010}. Importantly, theoretical investigations of photon dynamics and nonequilibrium critical behavior in driven light-matter systems suggest that anisotropy provides an additional degree of control over excitation pathways and multiphoton processes~\cite{yi-xiangGoldstoneHiggsModes2013,schiroPhaseTransitionLight2012a}.  
Therefore, the tunability of anisotropic interactions offers a promising route toward manipulating correlation properties and exploring richer QRM-related nonclassical photon statistics in experimentally relevant regimes.

For the QRSM, the Stark coupling does not need to be artificially introduced but can arise naturally or effectively in various QED systems. In the presence of strong driving and in the dispersive regime, significant Stark shifts emerge in cavity–atom systems, leading to effective Rabi–Stark–type models~\cite{Grimsmo2013PRA_87_033814,Grimsmo2014PRA_89_}. More recently, Zhai~\textit{et al.}  proposed a trapped-ion implementation in which tunable Stark interactions are engineered via laser-induced Stark shifts~\cite{Zhai2025PRA_112_013720}, providing a clear experimental pathway for exploiting Stark coupling to control multiphoton processes. 
Cong \emph{et al.} proposed a trapped-ion scheme in which the effective Stark term originates from laser-driven level shifts, while the anisotropy is tuned by independently adjusting the amplitudes of red- and blue-sideband drivings~\cite{congSelectiveInteractionsQuantum2020}.

In summary, the two-photon bundle control mechanisms proposed in this work are grounded in tunable parameters that are already accessible in strongly coupled light–matter systems. Key control knobs, including coupling strength, anisotropy, and Stark interaction, can be independently or jointly engineered across a variety of experimental platforms. This high degree of experimental accessibility renders the predicted photon-bundle dynamics directly amenable to experimental exploration and highlights the potential of these schemes for realizing tunable nonclassical multiphoton light sources. These considerations place our theoretical results within a physically complete framework and naturally connect to the conclusions drawn below.

\section{Conclusion}
\label{sec: Conclusion}
In this work, we have explored two-photon bundle correlations across the QRM family, e.g., QRM, AQRM, QRSM, and AQRSM, within the DME framework for open quantum systems. Our analysis demonstrates that distinct quantum light-matter couplings, anisotropy, and Stark terms offer versatile knobs to manipulate photon-bundle correlations at finite temperature, 
which enables controllable transitions between bunching and antibunching regimes.

Specifically, low-temperature QRM exhibits two successive dynamical transitions between bunching and antibunching, which can be exploited to realize two-photon bundle sources. In AQRM, anisotropic coupling enriches the correlation landscape: strong anisotropy enlarges antibunching regions, favoring two-photon bundle guns, while weak anisotropy creates a divergent response between single- and two-photon correlations at the first-excited-state phase transition, highlighting optimal conditions for two-photon bundle lasing and enabling prediction of excited-state QTPs. In QRSM, the Stark term acts as a decisive control: positive Stark coupling enhances bunching in regions with larger light–matter interaction strength, aligned with single-photon antibunching, whereas negative Stark coupling produces strong bunching at lower interaction strengths, along with additional antibunching regions suitable for pure two-photon guns. Finally, in AQRSM, anisotropy and Stark effects interplay to provide versatile control, with positive Stark coupling enhancing antibunching under strong anisotropy (favoring bundle guns) and negative Stark coupling amplifying bunching (promoting two-photon bundle lasing). Across all models, strong-bunching regions at $g>1$ consistently correlate with excited-state phase transitions, with complementary singularities in single-photon correlations serving as clear operational fingerprints for targeted quantum light generation.

Overall, our results provide a unified framework for engineering two-photon bundle sources, highlighting how systematic tuning of system parameters enables precise control over quantum light correlations, with potential applications in quantum optics~\cite{Liu2019NN_14_586-593,Wang2019NP_13_770-775} and quantum information processing ~\cite{Kok2007RMP_79_135-174,Raussendorf2007PRL_98_190504,Aaronson2011PFAASTC__333–342,Gambetta2017nQI_3_2,Liu2019NN_14_586-593,Wang2019NP_13_770-775,Cleve1997PRA_56_1201-1204,Braunstein2005RMP_77_513-577,Kimble2008N_453_1023-1030,Scarani2009RMP_81_1301-1350,Wang2019PRA_99_042309,Caves1980RMP_52_341-392b,Grangier1987PRL_59_2153-2156,Xiao1987PRL_59_278-281,Polzik1992PRL_68_3020-3023,Aasi2013NP_7_613-619,Pedrozo-Penafiel2020N_588_414-418,LIGOScientificCollaborationandVirgoCollaboration2021PRX_11_021053,Grote2013PRL_110_181101}. 


\begin{acknowledgments}
 This work is supported by the National Key R\&D Program of China under Grant No. 2024YFA1408900 (YXZ and QHC) and the Zhejiang Provincial Natural Science Foundation of China under Grant No. LZ25A050001 (CW).
\end{acknowledgments}


\bibliography{main}

@PREAMBLE{
 "\providecommand{\noopsort}[1]{}" 
 # "\providecommand{\singleletter}[1]{#1}%" 
}

@article{Chen2021PRA_103_043708,
  title = {Multiple Ground-State Instabilities in the Anisotropic Quantum {{Rabi}} Model},
  author = {Chen, Xiang-You and Duan, Liwei and Braak, Daniel and Chen, Qing-Hu},
  year = {2021},
  month = apr,
  journal = {Phys. Rev. A},
  volume = {103},
  number = {4},
  pages = {043708},
  doi = {10.1103/PhysRevA.103.043708},
  langid = {english}
}

@article{Liu2017PRL_119_,
  title = {Universal {{Scaling}} and {{Critical Exponents}} of the {{Anisotropic Quantum Rabi Model}}},
  author = {Liu, Maoxin and Chesi, Stefano and Ying, Zu-Jian and Chen, Xiaosong and Luo, Hong-Gang and Lin, Hai-Qing},
  year = {2017},
  month = nov,
  journal = {Phys. Rev. Lett.},
  volume = {119},
  number = {22},
  doi = {10.1103/physrevlett.119.220601},
  copyright = {https://link.aps.org/licenses/aps-default-license},
  langid = {english}
}

@article{Xie2014PRX_4_021046,
  title = {Anisotropic {{Rabi}} Model},
  author = {Xie, Qiong-Tao and Cui, Shuai and Cao, Jun-Peng and Amico, Luigi and Fan, Heng},
  year = {2014},
  month = jun,
  journal = {Phys. Rev. X},
  volume = {4},
  number = {2},
  pages = {021046},
  doi = {10.1103/PhysRevX.4.021046},
  copyright = {http://creativecommons.org/licenses/by/3.0/},
  langid = {english}
}

@article{Ye2023PSMIA_609_128364,
  title = {Implication of Giant Photon Bunching on Quantum Phase Transition in the Dissipative Anisotropic Quantum {{Rabi}} Model},
  author = {Ye, Tian and Wang, Chen and Chen, Qing-Hu},
  year = {2023},
  month = jan,
  journal = {Physica A: Statistical Mechanics and its Applications},
  volume = {609},
  pages = {128364},
  doi = {10.1016/j.physa.2022.128364},
  copyright = {https://www.elsevier.com/tdm/userlicense/1.0/},
  langid = {english}
}

@article{Ye2024OE_32_33483,
  title = {Anisotropic Qubit-Photon Interactions Inducing Multiple Antibunching-to-Bunching Transitions of Photons},
  author = {Ye, Tian and Wang, Chen and Chen, Qing-Hu},
  year = {2024},
  month = sep,
  journal = {Opt. Express},
  volume = {32},
  number = {19},
  pages = {33483},
  doi = {10.1364/OE.533310},
  langid = {english}
}

@article{Bienfait2019S__,
  title = {Phonon-Mediated Quantum State Transfer and Remote Qubit Entanglement},
  author = {Bienfait, A. and Satzinger, K. J. and Zhong, Y. P. and Chang, H.-S. and Chou, M.-H. and Conner, C. R. and Dumur, {\'E} and Grebel, J. and Peairs, G. A. and Povey, R. G. and Cleland, A. N.},
  year = {2019},
  month = apr,
  journal = {Science},
  doi = {10.1126/science.aaw8415},
  copyright = {Copyright {\copyright} 2019 The Authors, some rights reserved; exclusive licensee American Association for the Advancement of Science. No claim to original U.S. Government Works},
  langid = {english}
}

@article{Boto2000PRL_85_2733-2736,
  title = {Quantum {{Interferometric Optical Lithography}}: {{Exploiting Entanglement}} to {{Beat}} the {{Diffraction Limit}}},
  shorttitle = {Quantum {{Interferometric Optical Lithography}}},
  author = {Boto, Agedi N. and Kok, Pieter and Abrams, Daniel S. and Braunstein, Samuel L. and Williams, Colin P. and Dowling, Jonathan P.},
  year = {2000},
  month = sep,
  journal = {Phys. Rev. Lett.},
  volume = {85},
  number = {13},
  pages = {2733},
  doi = {10.1103/PhysRevLett.85.2733}
}

@article{congSelectiveInteractionsQuantum2020,
  title = {Selective Interactions in the Quantum {{Rabi}} Model},
  author = {Cong, L. and Felicetti, S. and Casanova, J. and Lamata, L. and Solano, E. and Arrazola, I.},
  year = {2020},
  month = mar,
  journal = {Phys. Rev. A},
  volume = {101},
  number = {3},
  pages = {032350},
  doi = {10.1103/physreva.101.032350},
  copyright = {https://link.aps.org/licenses/aps-default-license},
  langid = {english}
}

@article{xieFirstorderContinuousQuantum2020,
  title = {First-Order and Continuous Quantum Phase Transitions in the Anisotropic Quantum {{Rabi-Stark}} Model},
  author = {Xie, You-Fei and Chen, Xiang-You and Dong, Xiao-Fei and Chen, Qing-Hu},
  year = {2020},
  month = may,
  journal = {Phys. Rev. A},
  volume = {101},
  number = {5},
  pages = {053803},
  doi = {10.1103/PhysRevA.101.053803},
  langid = {english}
}

@article{xuExploringRoleCriticality2024,
  title = {Exploring the Role of Criticality in the Quantum {{Otto}} Cycle Fueled by the Anisotropic Quantum {{Rabi-Stark}} Model},
  author = "Xu, He-Guang and Jin, Jiasen and {de Almeida}, Norton G. and {de Moraes Neto}, G. D.",
  year = {2024},
  month = oct,
  journal = {Phys. Rev. B},
  volume = {110},
  number = {13},
  pages = {134318},
  doi = {10.1103/PhysRevB.110.134318},
  langid = {english}
}

@article{Braunstein2005RMP_77_513-577,
  title = {Quantum Information with Continuous Variables},
  author = {Braunstein, Samuel L. and {van Loock}, Peter},
  year = {2005},
  month = jun,
  journal = {Rev. Mod. Phys.},
  volume = {77},
  number = {2},
  pages = {513},
  doi = {10.1103/RevModPhys.77.513}
}

@article{Cleve1997PRA_56_1201-1204,
  title = {Substituting Quantum Entanglement for Communication},
  author = {Cleve, Richard and Buhrman, Harry},
  year = {1997},
  month = aug,
  journal = {Phys. Rev. A},
  volume = {56},
  number = {2},
  pages = {1201},
  doi = {10.1103/PhysRevA.56.1201}
}

@article{Kimble2008N_453_1023-1030,
  title = {The Quantum Internet},
  author = {Kimble, H. J.},
  year = {2008},
  month = jun,
  journal = {Nature},
  volume = {453},
  number = {7198},
  pages = {1023},
  doi = {10.1038/nature07127}
}

@article{Scarani2009RMP_81_1301-1350,
  title = {The Security of Practical Quantum Key Distribution},
  author = {Scarani, Valerio and {Bechmann-Pasquinucci}, Helle and Cerf, Nicolas J. and Du{\v s}ek, Miloslav and L{\"u}tkenhaus, Norbert and Peev, Momtchil},
  year = {2009},
  month = sep,
  journal = {Rev. Mod. Phys.},
  volume = {81},
  number = {3},
  pages = {1301},
  doi = {10.1103/RevModPhys.81.1301}
}

@article{Wang2019PRA_99_042309,
  title = {Continuous-Variable Measurement-Device-Independent Quantum Key Distribution Using Modulated Squeezed States and Optical Amplifiers},
  author = {Wang, Pu and Wang, Xuyang and Li, Yongmin},
  year = {2019},
  month = apr,
  journal = {Phys. Rev. A},
  volume = {99},
  number = {4},
  pages = {042309},
  doi = {10.1103/PhysRevA.99.042309}
}

@inproceedings{Aaronson2011PFAASTC__333–342,
  title = {The Computational Complexity of Linear Optics},
  booktitle = {Proceedings of the Forty-Third Annual {{ACM}} Symposium on {{Theory}} of Computing},
  author = {Aaronson, Scott and Arkhipov, Alex},
  year = {2011},
  month = jun,
  series = {{{STOC}} '11},
  pages = {333},
  address = {New York, NY, USA},
  doi = {10.1145/1993636.1993682},
  isbn = {978-1-4503-0691-1}
}

@article{Gambetta2017nQI_3_2,
  title = {Building Logical Qubits in a Superconducting Quantum Computing System},
  author = {Gambetta, Jay M. and Chow, Jerry M. and Steffen, Matthias},
  year = {2017},
  month = jan,
  journal = {npj Quantum Inf},
  volume = {3},
  number = {1},
  pages = {2},
  doi = {10.1038/s41534-016-0004-0},
  copyright = {2017 The Author(s)},
  langid = {english}
}

@article{Kok2007RMP_79_135-174,
  title = {Linear Optical Quantum Computing with Photonic Qubits},
  author = {Kok, Pieter and Munro, W. J. and Nemoto, Kae and Ralph, T. C. and Dowling, Jonathan P. and Milburn, G. J.},
  year = {2007},
  month = jan,
  journal = {Rev. Mod. Phys.},
  volume = {79},
  number = {1},
  pages = {135},
  doi = {10.1103/RevModPhys.79.135}
}

@article{Liu2019NN_14_586-593,
  title = {A Solid-State Source of Strongly Entangled Photon Pairs with High Brightness and Indistinguishability},
  author = {Liu, Jin and Su, Rongbin and Wei, Yuming and Yao, Beimeng and da Silva, Saimon Filipe Covre and Yu, Ying and {Iles-Smith}, Jake and Srinivasan, Kartik and Rastelli, Armando and Li, Juntao and Wang, Xuehua},
  year = {2019},
  month = jun,
  journal = {Nat. Nanotechnol.},
  volume = {14},
  number = {6},
  pages = {586},
  doi = {10.1038/s41565-019-0435-9},
  copyright = {2019 The Author(s), under exclusive licence to Springer Nature Limited},
  langid = {english}
}

@article{Raussendorf2007PRL_98_190504,
  title = {Fault-{{Tolerant Quantum Computation}} with {{High Threshold}} in {{Two Dimensions}}},
  author = {Raussendorf, Robert and Harrington, Jim},
  year = {2007},
  month = may,
  journal = {Phys. Rev. Lett.},
  volume = {98},
  number = {19},
  pages = {190504},
  doi = {10.1103/PhysRevLett.98.190504}
}

@article{Wang2019NP_13_770-775,
  title = {Towards Optimal Single-Photon Sources from Polarized Microcavities},
  author = {Wang, Hui and He, Yu-Ming and Chung, T.-H. and Hu, Hai and Yu, Ying and Chen, Si and Ding, Xing and Chen, M.-C. and Qin, Jian and Yang, Xiaoxia and Liu, Run-Ze and Duan, Z.-C. and Li, J.-P. and Gerhardt, S. and Winkler, K. and Jurkat, J. and Wang, Lin-Jun and Gregersen, Niels and Huo, Yong-Heng and Dai, Qing and Yu, Siyuan and H{\"o}fling, Sven and Lu, Chao-Yang and Pan, Jian-Wei},
  year = {2019},
  month = nov,
  journal = {Nat. Photonics},
  volume = {13},
  number = {11},
  pages = {770},
  doi = {10.1038/s41566-019-0494-3},
  copyright = {2019 The Author(s), under exclusive licence to Springer Nature Limited},
  langid = {english}
}

@article{Beaudoin2011PRA_84_043832,
  title = {Dissipation and Ultrastrong Coupling in Circuit {{QED}}},
  author = {Beaudoin, F{\'e}lix and Gambetta, Jay M. and Blais, A.},
  year = {2011},
  month = oct,
  journal = {Phys. Rev. A},
  volume = {84},
  number = {4},
  pages = {043832},
  doi = {10.1103/PhysRevA.84.043832}
}

@article{Ridolfo2012PRL_109_193602,
  title = {Photon {{Blockade}} in the {{Ultrastrong Coupling Regime}}},
  author = {Ridolfo, A. and Leib, M. and Savasta, S. and Hartmann, M. J.},
  year = {2012},
  month = nov,
  journal = {Phys. Rev. Lett.},
  volume = {109},
  number = {19},
  pages = {193602},
  doi = {10.1103/PhysRevLett.109.193602}
}

@article{Bin2020PRL_124_053601,
  title = {\${{N}}\$-{{Phonon Bundle Emission}} via the {{Stokes Process}}},
  author = {Bin, Qian and L{\"u}, Xin-You and Laussy, Fabrice P. and Nori, Franco and Wu, Ying},
  year = {2020},
  month = feb,
  journal = {Phys. Rev. Lett.},
  volume = {124},
  number = {5},
  pages = {053601},
  doi = {10.1103/PhysRevLett.124.053601}
}

@article{Bin2021PRL_127_073602,
  title = {Parity-{{Symmetry-Protected Multiphoton Bundle Emission}}},
  author = {Bin, Qian and Wu, Ying and L{\"u}, Xin-You},
  year = {2021},
  month = aug,
  journal = {Phys. Rev. Lett.},
  volume = {127},
  number = {7},
  pages = {073602},
  doi = {10.1103/PhysRevLett.127.073602}
}

@article{Munoz2014NP_8_550-555,
  title = {Emitters of {{N-photon}} Bundles},
  author = {Mu{\~n}oz, C. S{\'a}nchez and {del Valle}, E. and Tudela, A. Gonz{\'a}lez and M{\"u}ller, K. and Lichtmannecker, S. and Kaniber, M. and Tejedor, C. and Finley, J. J. and Laussy, F. P.},
  year = {2014},
  month = jul,
  journal = {Nature Photon},
  volume = {8},
  number = {7},
  pages = {550},
  doi = {10.1038/nphoton.2014.114},
  copyright = {2014 Springer Nature Limited},
  langid = {english}
}

@article{Munoz2018OO_5_14-26,
  title = {Filtering Multiphoton Emission from State-of-the-Art Cavity Quantum Electrodynamics},
  author = {Mu{\~n}oz, Carlos S{\'a}nchez and Laussy, Fabrice P. and del Valle, Elena and Tejedor, Carlos and {Gonz{\'a}lez-Tudela}, Alejandro},
  year = {2018},
  month = jan,
  journal = {Optica, OPTICA},
  volume = {5},
  number = {1},
  pages = {14},
  doi = {10.1364/OPTICA.5.000014},
  copyright = {{\copyright} 2018 Optical Society of America},
  langid = {english}
}

@article{BROWN1956N_177_27-29,
  title = {Correlation between {{Photons}} in Two {{Coherent Beams}} of {{Light}}},
  author = {Hanbury Brown, R and Twiss, R Q},
  year = {1956},
  month = jan,
  journal = {Nature},
  volume = {177},
  number = {4497},
  pages = {27},
  doi = {10.1038/177027a0}
}

@article{Glauber1963PR_130_2529-2539,
  title = {The {{Quantum Theory}} of {{Optical Coherence}}},
  author = {Glauber, Roy J.},
  year = {1963},
  month = jun,
  journal = {Phys. Rev.},
  volume = {130},
  number = {6},
  pages = {2529},
  doi = {10.1103/physrev.130.2529},
  copyright = {http://link.aps.org/licenses/aps-default-license},
  langid = {english}
}

@article{Genoni2015NJP_17_013034,
  title = {Squeezing of Mechanical Motion via Qubit-Assisted Control},
  author = {Genoni, Marco G and Bina, Matteo and Olivares, Stefano and De Chiara, Gabriele and Paternostro, Mauro},
  year = {2015},
  month = jan,
  journal = {New J. Phys.},
  volume = {17},
  number = {1},
  pages = {013034},
  doi = {10.1088/1367-2630/17/1/013034},
  copyright = {http://iopscience.iop.org/info/page/text-and-data-mining},
  langid = {english}
}

@book{Weiss2021__,
author = {Weiss, Ulrich},
title = {Quantum Dissipative Systems},
publisher = {World Scientific},
year = {2012},
doi = {10.1142/8334},
address = {},
edition   = {4th},
URL = {https://www.worldscientific.com/doi/abs/10.1142/8334},
eprint = {https://www.worldscientific.com/doi/pdf/10.1142/8334}
}

@article{Aasi2013NP_7_613-619,
  title = {Enhanced Sensitivity of the {{LIGO}} Gravitational Wave Detector by Using Squeezed States of Light},
  author = {Aasi, J. and Abadie, J. and Abbott, B. P. and Abbott, R. and Abbott, T. D. and Abernathy, M. R. and Adams, C. and Adams, T. and Addesso, P. and Adhikari, R. X. and Affeldt, C. and Aguiar, O. D. and Ajith, P. and Allen, B. and Amador Ceron, E. and Amariutei, D. and Anderson, S. B. and Anderson, W. G. and Arai, K. and Araya, M. C. and Arceneaux, C. and Ast, S. and Aston, S. M. and Atkinson, D. and Aufmuth, P. and Aulbert, C. and Austin, L. and Aylott, B. E. and Babak, S. and Baker, P. T. and Ballmer, S. and Bao, Y. and Barayoga, J. C. and Barker, D. and Barr, B. and Barsotti, L. and Barton, M. A. and Bartos, I.},
  year = {2013},
  month = aug,
  journal = {Nature Photon},
  volume = {7},
  number = {8},
  pages = {613},
  doi = {10.1038/nphoton.2013.177},
  copyright = {2013 Springer Nature Limited},
  langid = {english}
}

@article{Caves1980RMP_52_341-392b,
  title = {On the Measurement of a Weak Classical Force Coupled to a Quantum-Mechanical Oscillator. {{I}}. {{Issues}} of Principle},
  author = {Caves, Carlton M. and Thorne, Kip S. and Drever, Ronald W. P. and Sandberg, Vernon D. and Zimmermann, Mark},
  year = {1980},
  month = apr,
  journal = {Rev. Mod. Phys.},
  volume = {52},
  number = {2},
  pages = {341},
  doi = {10.1103/RevModPhys.52.341}
}

@article{Colombo2022NP_18_925-930,
  title = {Time-Reversal-Based Quantum Metrology with Many-Body Entangled States},
  author = {Colombo, Simone and {Pedrozo-Pe{\~n}afiel}, Edwin and Adiyatullin, Albert F. and Li, Zeyang and Mendez, Enrique and Shu, Chi and Vuleti{\'c}, Vladan},
  year = {2022},
  month = aug,
  journal = {Nat. Phys.},
  volume = {18},
  number = {8},
  pages = {925},
  doi = {10.1038/s41567-022-01653-5},
  copyright = {2022 The Author(s), under exclusive licence to Springer Nature Limited},
  langid = {english}
}

@article{Grangier1987PRL_59_2153-2156,
  title = {Squeezed-Light--Enhanced Polarization Interferometer},
  author = {Grangier, P. and Slusher, R. E. and Yurke, B. and LaPorta, A.},
  year = {1987},
  month = nov,
  journal = {Phys. Rev. Lett.},
  volume = {59},
  number = {19},
  pages = {2153},
  doi = {10.1103/PhysRevLett.59.2153}
}

@article{Grote2013PRL_110_181101,
  title = {First {{Long-Term Application}} of {{Squeezed States}} of {{Light}} in a {{Gravitational-Wave Observatory}}},
  author = {Grote, H. and Danzmann, K. and Dooley, K. L. and Schnabel, R. and Slutsky, J. and Vahlbruch, H.},
  year = {2013},
  month = may,
  journal = {Phys. Rev. Lett.},
  volume = {110},
  number = {18},
  pages = {181101},
  doi = {10.1103/PhysRevLett.110.181101}
}

@article{LIGOScientificCollaborationandVirgoCollaboration2021PRX_11_021053,
  title = {{{GWTC-2}}: {{Compact Binary Coalescences Observed}} by {{LIGO}} and {{Virgo}} during the {{First Half}} of the {{Third Observing Run}}},
  shorttitle = {{{GWTC-2}}},
  author = {Abbott, R. and Abbott, T. D. and Abraham, S. and Acernese, F. and Ackley, K. and Adams, A. and Adams, C. and Adhikari, R. X. and Adya, V. B. and Affeldt, C. and Agathos, M. and Agatsuma, K. and Aggarwal, N. and Aguiar, O. D. and Aiello, L. and Ain, A. and Ajith, P. and Akcay, S. and Allen, G. and Allocca, A. and Altin, P. A. and Amato, A. and Anand, S. and Ananyeva, A. and Anderson, S. B. and Anderson, W. G. and Angelova, S. V. and Ansoldi, S. and Antelis, J. M. and Antier, S.},
  year = {2021},
  month = jun,
  journal = {Phys. Rev. X},
  volume = {11},
  number = {2},
  pages = {021053},
  doi = {10.1103/PhysRevX.11.021053}
}

@article{Moller2017N_547_191-195,
  title = {Quantum Back-Action-Evading Measurement of Motion in a Negative Mass Reference Frame},
  author = {M{\o}ller, Christoffer B. and Thomas, Rodrigo A. and Vasilakis, Georgios and Zeuthen, Emil and Tsaturyan, Yeghishe and Balabas, Mikhail and Jensen, Kasper and Schliesser, Albert and Hammerer, Klemens and Polzik, Eugene S.},
  year = {2017},
  month = jul,
  journal = {Nature},
  volume = {547},
  number = {7662},
  pages = {191},
  doi = {10.1038/nature22980}
}

@article{Motes2015PRL_114_170802,
  title = {Linear {{Optical Quantum Metrology}} with {{Single Photons}}: {{Exploiting Spontaneously Generated Entanglement}} to {{Beat}} the {{Shot-Noise Limit}}},
  shorttitle = {Linear {{Optical Quantum Metrology}} with {{Single Photons}}},
  author = {Motes, Keith R. and Olson, Jonathan P. and Rabeaux, Evan J. and Dowling, Jonathan P. and Olson, S. Jay and Rohde, Peter P.},
  year = {2015},
  month = apr,
  journal = {Phys. Rev. Lett.},
  volume = {114},
  number = {17},
  pages = {170802},
  doi = {10.1103/PhysRevLett.114.170802}
}

@article{Pedrozo-Penafiel2020N_588_414-418,
  title = {Entanglement on an Optical Atomic-Clock Transition},
  author = {{Pedrozo-Pe{\~n}afiel}, Edwin and Colombo, Simone and Shu, Chi and Adiyatullin, Albert F. and Li, Zeyang and Mendez, Enrique and Braverman, Boris and Kawasaki, Akio and Akamatsu, Daisuke and Xiao, Yanhong and Vuleti{\'c}, Vladan},
  year = {2020},
  month = dec,
  journal = {Nature},
  volume = {588},
  number = {7838},
  pages = {414},
  doi = {10.1038/s41586-020-3006-1},
  copyright = {2020 The Author(s), under exclusive licence to Springer Nature Limited},
  langid = {english}
}

@article{Polzik1992PRL_68_3020-3023,
  title = {Spectroscopy with Squeezed Light},
  author = {Polzik, E. S. and Carri, J. and Kimble, H. J.},
  year = {1992},
  month = may,
  journal = {Phys. Rev. Lett.},
  volume = {68},
  number = {20},
  pages = {3020},
  doi = {10.1103/PhysRevLett.68.3020}
}

@article{Qin2023PRL_130_070801,
  title = {Unconditional and {{Robust Quantum Metrological Advantage}} beyond {{N00N States}}},
  author = {Qin, Jian and Deng, Yu-Hao and Zhong, Han-Sen and Peng, Li-Chao and Su, Hao and Luo, Yi-Han and Xu, Jia-Min and Wu, Dian and Gong, Si-Qiu and Liu, Hua-Liang and Wang, Hui and Chen, Ming-Cheng and Li, Li and Liu, Nai-Le and Lu, Chao-Yang and Pan, Jian-Wei},
  year = {2023},
  month = feb,
  journal = {Phys. Rev. Lett.},
  volume = {130},
  number = {7},
  pages = {070801},
  doi = {10.1103/PhysRevLett.130.070801}
}

@article{Xiao1987PRL_59_278-281,
  title = {Precision Measurement beyond the Shot-Noise Limit},
  author = {Xiao, Min and Wu, Ling-An and Kimble, H. J.},
  year = {1987},
  month = jul,
  journal = {Phys. Rev. Lett.},
  volume = {59},
  number = {3},
  pages = {278},
  doi = {10.1103/PhysRevLett.59.278}
}

@article{Braak2016JPAMT_49_300301,
  title = {Semi-Classical and Quantum {{Rabi}} Models: In Celebration of 80 Years},
  shorttitle = {Semi-Classical and Quantum {{Rabi}} Models},
  author = {Braak, Daniel and Chen, Qing-Hu and Batchelor, Murray T and Solano, Enrique},
  year = {2016},
  month = jun,
  journal = {J. Phys. A: Math. Theor.},
  volume = {49},
  number = {30},
  pages = {300301},
  doi = {10.1088/1751-8113/49/30/300301},
  langid = {english}
}

@article{Casanova2010PRL_105_263603,
  title = {Deep {{Strong Coupling Regime}} of the {{Jaynes-Cummings Model}}},
  author = {Casanova, J. and Romero, G. and Lizuain, I. and {Garc{\'i}a-Ripoll}, J. J. and Solano, E.},
  year = {2010},
  month = dec,
  journal = {Phys. Rev. Lett.},
  volume = {105},
  number = {26},
  pages = {263603},
  doi = {10.1103/PhysRevLett.105.263603}
}

@article{Chen2022JPBAMOP_55_115502,
  title = {Nonclassical Photon Statistics and Photon Squeezing in the Dissipative Mixed Quantum {{Rabi}} Model},
  author = {Chen, Xu-Min and Chen, Zhe-Kai and Che, Han-Xin and Wang, Chen},
  year = {2022},
  month = jun,
  journal = {J. Phys. B: At. Mol. Opt. Phys.},
  volume = {55},
  number = {11},
  pages = {115502},
  doi = {10.1088/1361-6455/ac6bd5},
  copyright = {https://iopscience.iop.org/page/copyright},
  langid = {english}
}

@article{Clarke2008N_453_1031-1042,
  title = {Superconducting Quantum Bits},
  author = {Clarke, John and Wilhelm, Frank K.},
  year = {2008},
  month = jun,
  journal = {Nature},
  volume = {453},
  number = {7198},
  pages = {1031},
  doi = {10.1038/nature07128},
  copyright = {2007 Springer Nature Limited},
  langid = {english}
}

@article{Forn-Diaz2010PRL_105_237001,
  title = {Observation of the {{Bloch-Siegert Shift}} in a {{Qubit-Oscillator System}} in the {{Ultrastrong Coupling Regime}}},
  author = {{Forn-D{\'i}az}, P. and Lisenfeld, J. and Marcos, D. and {Garc{\'i}a-Ripoll}, J. J. and Solano, E. and Harmans, C. J. P. M. and Mooij, J. E.},
  year = {2010},
  month = nov,
  journal = {Phys. Rev. Lett.},
  volume = {105},
  number = {23},
  pages = {237001},
  doi = {10.1103/PhysRevLett.105.237001}
}

@article{Rabi1936PR_49_324-328,
  title = {On the {{Process}} of {{Space Quantization}}},
  author = {Rabi, I. I.},
  year = {1936},
  month = feb,
  journal = {Phys. Rev.},
  volume = {49},
  number = {4},
  pages = {324},
  doi = {10.1103/PhysRev.49.324}
}

@article{Ridolfo2013PRL_110_163601,
  title = {Nonclassical {{Radiation}} from {{Thermal Cavities}} in the {{Ultrastrong Coupling Regime}}},
  author = {Ridolfo, A. and Savasta, S. and Hartmann, M. J.},
  year = {2013},
  month = apr,
  journal = {Phys. Rev. Lett.},
  volume = {110},
  number = {16},
  pages = {163601},
  doi = {10.1103/PhysRevLett.110.163601}
}

@book{Scully1997__,
  title = {Quantum {{Optics}}},
  author = {Scully, Marlan O. and Zubairy, M. Suhail},
  year = {1997},
  address = {Cambridge},
 publisher = {Cambridge University Press},
  doi = {10.1017/CBO9780511813993},
  isbn = {978-0-521-43595-6}
}

@article{Chen2020PRA_102_,
  title = {Quantum Criticality of the {{Rabi-Stark}} Model at Finite Frequency Ratios},
  author = {Chen, Xiang-You and Xie, You-Fei and Chen, Qing-Hu},
  year = {2020},
  month = dec,
  journal = {Phys. Rev. A},
  volume = {102},
  number = {6},
  pages = {063721},
  doi = {10.1103/physreva.102.063721},
  copyright = {https://link.aps.org/licenses/aps-default-license},
  langid = {english}
}

@article{Eckle2017JPAMT_50_294004,
  title = {A Generalization of the Quantum {{Rabi}} Model: Exact Solution and Spectral Structure},
  shorttitle = {A Generalization of the Quantum {{Rabi}} Model},
  author = {Eckle, Hans-Peter and Johannesson, Henrik},
  year = {2017},
  month = jul,
  journal = {J. Phys. A: Math. Theor.},
  volume = {50},
  number = {29},
  pages = {294004},
  doi = {10.1088/1751-8121/aa785a},
  copyright = {http://iopscience.iop.org/info/page/text-and-data-mining},
  langid = {english}
}

@article{Grimsmo2013PRA_87_033814,
  title = {Cavity-{{QED}} Simulation of Qubit-Oscillator Dynamics in the Ultrastrong-Coupling Regime},
  author = {Grimsmo, Arne L. and Parkins, Scott},
  year = {2013},
  month = mar,
  journal = {Phys. Rev. A},
  volume = {87},
  number = {3},
  pages = {033814},
  doi = {10.1103/PhysRevA.87.033814}
}

@article{Xie2019JPAMT_52_245304,
  title = {Quantum {{Rabi}}--{{Stark}} Model: Solutions and Exotic Energy Spectra},
  shorttitle = {Quantum {{Rabi}}--{{Stark}} Model},
  author = {Xie, You-Fei and Duan, Liwei and Chen, Qing-Hu},
  year = {2019},
  month = jun,
  journal = {J. Phys. A: Math. Theor.},
  volume = {52},
  number = {24},
  pages = {245304},
  doi = {10.1088/1751-8121/ab1cf6},
  copyright = {http://iopscience.iop.org/info/page/text-and-data-mining},
  langid = {english}
}

@article{Grimsmo2014PRA_89_,
  title = {Open {{Rabi}} Model with Ultrastrong Coupling plus Large Dispersive-Type Nonlinearity: {{Nonclassical}} Light via a Tailored Degeneracy},
  shorttitle = {Open {{Rabi}} Model with Ultrastrong Coupling plus Large Dispersive-Type Nonlinearity},
  author = {Grimsmo, Arne L.},
  year = {2014},
  journal = {Phys. Rev. A},
  volume = {89},
  number = {3},
  doi = {10.1103/PhysRevA.89.033802}
}

@article{Chang2016PRL_117_203602a,
  title = {Deterministic {{Down-Converter}} and {{Continuous Photon-Pair Source}} within the {{Bad-Cavity Limit}}},
  author = {Chang, Yue and {Gonz{\'a}lez-Tudela}, Alejandro and S{\'a}nchez Mu{\~n}oz, Carlos and {Navarrete-Benlloch}, Carlos and Shi, Tao},
  year = {2016},
  month = {nov},
  journal = {Phys. Rev. Lett.},
  volume = {117},
  number = {20},
  pages = {203602},
  doi = {10.1103/PhysRevLett.117.203602},
  copyright = {http://link.aps.org/licenses/aps-default-license},
  langid = {english}
}

@article{Jiang2023OE_31_15697,
  title = {Multiple-Photon Bundle Emission in the n-Photon {{Jaynes-Cummings}} Model},
  author = {Jiang, Shu-Yuan and Zou, Fen and Wang, Yi and Huang, Jin-Feng and Xu, Xun-Wei and Liao, Jie-Qiao},
  year = {2023},
  month = {may},
  journal = {Opt. Express},
  volume = {31},
  number = {10},
  pages = {15697},
  doi = {10.1364/OE.488167},
  langid = {english}
}

@article{Qin2019PRA_100_062501,
  title = {Emission of Photon Pairs by Mechanical Stimulation of the Squeezed Vacuum},
  author = {Qin, Wei and Macr{\`i}, Vincenzo and Miranowicz, Adam and Savasta, Salvatore and Nori, Franco},
  year = {2019},
  month = {dec},
  journal = {Phys. Rev. A},
  volume = {100},
  number = {6},
  pages = {062501},
  doi = {10.1103/PhysRevA.100.062501},
  langid = {english}
}

@article{Strekalov2014NP_8_500-501,
  title = {A Bundle of Photons, Please},
  author = {Strekalov, Dmitry V.},
  year = {2014},
  month = {jul},
  journal = {Nature Photon},
  volume = {8},
  number = {7},
  pages = {500--501},
  doi = {10.1038/nphoton.2014.144},
  copyright = {2014 Springer Nature Limited},
  langid = {english}
}

@article{Xiong2025PRR_7_013238,
  title = {Two-Photon Decay Enhanced Even Photon Bundle Emission},
  author = {Xiong, Biao and Bin, Qian and Chao, Shi-Lei and Liu, Ji-Bing and L{\"u}, Xin-You},
  year = {2025},
  month = {mar},
  journal = {Phys. Rev. Research},
  volume = {7},
  number = {1},
  pages = {013238},
  doi = {10.1103/PhysRevResearch.7.013238},
  langid = {english}
}

@article{dangeloTwoPhotonDiffractionQuantum2001,
  title = {Two-{{Photon Diffraction}} and {{Quantum Lithography}}},
  author = {D'Angelo, Milena and Chekhova, Maria V. and Shih, Yanhua},
  date = {2001-06-14},
  year = {2001},
  journal = {Phys. Rev. Lett.},
  volume = {87},
  number = {1},
  pages = {013602},
  publisher = {American Physical Society},
  doi = {10.1103/PhysRevLett.87.013602},
  url = {https://link.aps.org/doi/10.1103/PhysRevLett.87.013602},
  urldate = {2026-01-08}
}

@article{denkTwoPhotonLaserScanning1990a,
  title = {Two-{{Photon Laser Scanning Fluorescence Microscopy}}},
  author = {Denk, Winfried and Strickler, James H. and Webb, Watt W.},
  date = {1990-04-06},
  year = {1990},
  journal = {Science},
  volume = {248},
  number = {4951},
  pages = {73--76},
  publisher = {American Association for the Advancement of Science},
  doi = {10.1126/science.2321027},
  url = {https://www.science.org/doi/10.1126/science.2321027},
  urldate = {2026-01-08}
}

@article{dorfmanNonlinearOpticalSignals2016,
  title = {Nonlinear Optical Signals and Spectroscopy with Quantum Light},
  author = {Dorfman, Konstantin E. and Schlawin, Frank and Mukamel, Shaul},
  date = {2016-12-28},
  year = {2016},
  journal = {Rev. Mod. Phys.},
  volume = {88},
  number = {4},
  pages = {045008},
  publisher = {American Physical Society},
  doi = {10.1103/RevModPhys.88.045008},
  url = {https://link.aps.org/doi/10.1103/RevModPhys.88.045008},
  urldate = {2026-01-08}
}

@article{hortonVivoThreephotonMicroscopy2013,
  title = {In Vivo Three-Photon Microscopy of Subcortical Structures within an Intact Mouse Brain},
  author = {Horton, Nicholas G. and Wang, Ke and Kobat, Demirhan and Clark, Catharine G. and Wise, Frank W. and Schaffer, Chris B. and Xu, Chris},
  date = {2013-03},
  year = {2013},
  journal = {Nature Photon},
  volume = {7},
  number = {3},
  pages = {205--209},
  publisher = {Nature Publishing Group},
  issn = {1749-4893},
  doi = {10.1038/nphoton.2012.336},
  url = {https://www.nature.com/articles/nphoton.2012.336},
  urldate = {2026-01-08},
  langid = {english}
}

@article{liTwophotonNanoprobesBased2023,
  title = {Two-Photon Nanoprobes Based on Bioorganic Nanoarchitectonics with a Photo-Oxidation Enhanced Emission Mechanism},
  author = {Li, Shukun and Chang, Rui and Zhao, Luyang and Xing, Ruirui and van Hest, Jan C. M. and Yan, Xuehai},
  date = {2023-08-26},
  year = {2023},
  journal = {Nat Commun},
  volume = {14},
  number = {1},
  pages = {5227},
  publisher = {Nature Publishing Group},
  issn = {2041-1723},
  doi = {10.1038/s41467-023-40897-4},
  url = {https://www.nature.com/articles/s41467-023-40897-4},
  urldate = {2026-01-08},
  langid = {english}
}

@article{lopezcarrenoExcitingPolaritonsQuantum2015,
  title = {Exciting {{Polaritons}} with {{Quantum Light}}},
  author = {López Carreño, J. C. and Sánchez Muñoz, C. and Sanvitto, D. and del Valle, E. and Laussy, F. P.},
  date = {2015-11-04},
  year = {2015},
  journal = {Phys. Rev. Lett.},
  volume = {115},
  number = {19},
  pages = {196402},
  publisher = {American Physical Society},
  doi = {10.1103/PhysRevLett.115.196402},
  url = {https://link.aps.org/doi/10.1103/PhysRevLett.115.196402},
  urldate = {2026-01-08}
}

@article{xingDeterministicGenerationArbitrary2024a,
  title = {Deterministic Generation of Arbitrary \$n\$-Photon States in a Waveguide-{{QED}} System},
  author = {Xing, Fan and Liao, Zeyang and Wang, Xue-hua},
  date = {2024-01-22},
  year = {2024},
  journal = {Phys. Rev. A},
  volume = {109},
  number = {1},
  pages = {013718},
  publisher = {American Physical Society},
  doi = {10.1103/PhysRevA.109.013718},
  url = {https://link.aps.org/doi/10.1103/PhysRevA.109.013718},
  urldate = {2026-01-08}
}

@article{zouDynamicalEmissionPhonon2022b,
  title = {Dynamical Emission of Phonon Pairs in Optomechanical Systems},
  author = {Zou, Fen and Liao, Jie-Qiao and Li, Yong},
  date = {2022-05-10},
  year = {2022},
  journal = {Phys. Rev. A},
  volume = {105},
  number = {5},
  pages = {053507},
  issn = {2469-9926, 2469-9934},
  doi = {10.1103/PhysRevA.105.053507},
  url = {https://link.aps.org/doi/10.1103/PhysRevA.105.053507},
  urldate = {2026-01-08},
  langid = {english}
}

@article{bieniasScatteringResonancesBound2014,
  title = {Scattering Resonances and Bound States for Strongly Interacting {{Rydberg}} Polaritons},
  author = {Bienias, P. and Choi, S. and Firstenberg, O. and Maghrebi, M. F. and Gullans, M. and Lukin, M. D. and Gorshkov, A. V. and B{\"u}chler, H. P.},
  year = {2014},
  month = {nov},
  journal = {Phys. Rev. A},
  series = {Rydberg},
  volume = {90},
  number = {5},
  pages = {053804},
  issn = {1050-2947, 1094-1622},
  doi = {10.1103/PhysRevA.90.053804}
}

@article{douglasPhotonMoleculesAtomic2016,
  title = {Photon {{Molecules}} in {{Atomic Gases Trapped Near Photonic Crystal Waveguides}}},
  author = {Douglas, James S. and Caneva, Tommaso and Chang, Darrick E.},
  year = {2016},
  month = {aug},
  journal = {Phys. Rev. X},
  volume = {6},
  number = {3},
  pages = {031017},
  publisher = {American Physical Society},
  doi = {10.1103/PhysRevX.6.031017},
  urldate = {2026-01-08}
}

@article{gonzalez-tudelaEfficientMultiphotonGeneration2017a,
  title = {Efficient {{Multiphoton Generation}} in {{Waveguide Quantum Electrodynamics}}},
  author = {{Gonz{\'a}lez-Tudela}, A. and Paulisch, V. and Kimble, H. J. and Cirac, J. I.},
  year = {2017},
  month = {may},
  journal = {Phys. Rev. Lett.},
  volume = {118},
  number = {21},
  pages = {213601},
  publisher = {American Physical Society},
  doi = {10.1103/PhysRevLett.118.213601},
  urldate = {2026-01-08}
}

@article{jachymskiThreeBodyInteractionRydberg2016,
  title = {Three-{{Body Interaction}} of {{Rydberg Slow-Light Polaritons}}},
  author = {Jachymski, Krzysztof and Bienias, Przemys{\l}aw and B{\"u}chler, Hans Peter},
  year = {2016},
  month = {jul},
  journal = {Phys. Rev. Lett.},
  volume = {117},
  number = {5},
  pages = {053601},
  publisher = {American Physical Society},
  doi = {10.1103/PhysRevLett.117.053601},
  urldate = {2026-01-08}
}

@article{liangObservationThreephotonBound2018,
  title = {Observation of Three-Photon Bound States in a Quantum Nonlinear Medium},
  author = {Liang, Qi-Yu and Venkatramani, Aditya V. and Cantu, Sergio H. and Nicholson, Travis L. and Gullans, Michael J. and Gorshkov, Alexey V. and Thompson, Jeff D. and Chin, Cheng and Lukin, Mikhail D. and Vuleti{\'c}, Vladan},
  year = {2018},
  month = {feb},
  journal = {Science},
  volume = {359},
  number = {6377},
  pages = {783--786},
  publisher = {American Association for the Advancement of Science},
  doi = {10.1126/science.aao7293},
  urldate = {2026-01-08}
}

@article{liaoCorrelatedTwophotonScattering2013,
  title = {Correlated Two-Photon Scattering in Cavity Optomechanics},
  author = {Liao, Jie-Qiao and Law, C. K.},
  year = {2013},
  month = {apr},
  journal = {Phys. Rev. A},
  volume = {87},
  number = {4},
  pages = {043809},
  publisher = {American Physical Society},
  doi = {10.1103/PhysRevA.87.043809}
}

@article{liaoCorrelatedTwophotonTransport2010,
  title = {Correlated Two-Photon Transport in a One-Dimensional Waveguide Side-Coupled to a Nonlinear Cavity},
  author = {Liao, Jie-Qiao and Law, C. K.},
  year = {2010},
  month = {nov},
  journal = {Phys. Rev. A},
  volume = {82},
  number = {5},
  pages = {053836},
  issn = {1050-2947, 1094-1622},
  doi = {10.1103/PhysRevA.82.053836}
}

@article{Braak2024JOSAB_41_C97,
  title = {Spectral Continuum in the {{Rabi}}--{{Stark}} Model [{{Invited}}]},
  author = {Braak, Daniel and Cong, Lei and Eckle, Hans-Peter and Johannesson, Henrik and Twyeffort, Elinor K.},
  year = 2024,
  month = aug,
  journal = {J. Opt. Soc. Am. B},
  volume = {41},
  number = {8},
  pages = {C97},
  doi = {10.1364/JOSAB.524014},
  langid = {english}
}

@article{Zhai2025PRA_112_013720,
  title = {Stark-Induced Tunable Phase Transition in the Two-Photon {{Dicke-Stark}} Model},
  author = {Zhai, Cui-Lu and Wu, Wei and Wu, Chun-Wang and Chen, Ping-Xing},
  year = 2025,
  month = jul,
  journal = {Phys. Rev. A},
  volume = {112},
  number = {1},
  pages = {013720},
  doi = {10.1103/vqnf-zd5w},
  langid = {english}
}

@article{Ying2023AQT_6_2200068,
  title = {Scaling {{Relations}} and {{Topological Quadruple Points}} in {{Light-Matter Interactions}} with {{Anisotropy}} and {{Nonlinear Stark Coupling}}},
  author = {Ying, Zu-Jian},
  year = 2023,
  journal = {Advanced Quantum Technologies},
  volume = {6},
  number = {1},
  pages = {2200068},
  doi = {10.1002/qute.202200068},
  langid = {english}
}

@article{zhangNonclassicalCorrelationsQuadrature,
  title = {Nonclassical {{Correlations}} and {{Quadrature Squeezing}} of {{Photons}} in {{Anisotropic Quantum Rabi-Stark Model}}},
  author = {Zhang, Yong-Xin and Wang, Chen and Chen, Qing-Hu},
  journal = {Advanced Quantum Technologies},
  year = 2025,
  volume = {n/a},
  number = {n/a},
  pages = {e00744},
  issn = {2511-9044},
  doi = {10.1002/qute.202500744},
  urldate = {2026-01-08}
}

@article{englundControllingCavityReflectivity2007,
  title = {Controlling Cavity Reflectivity with a Single Quantum Dot},
  author = {Englund, Dirk and Faraon, Andrei and Fushman, Ilya and Stoltz, Nick and Petroff, Pierre and Vu{\v c}kovi{\'c}, Jelena},
  year = 2007,
  month = dec,
  journal = {Nature},
  volume = {450},
  number = {7171},
  pages = {857--861},
  publisher = {Nature Publishing Group},
  issn = {1476-4687},
  doi = {10.1038/nature06234}
}

@article{forn-diazUltrastrongCouplingRegimes2019,
  title = {Ultrastrong Coupling Regimes of Light-Matter Interaction},
  author = {{Forn-D{\'i}az}, P. and Lamata, L. and Rico, E. and Kono, J. and Solano, E.},
  year = 2019,
  month = jun,
  journal = {Rev. Mod. Phys.},
  volume = {91},
  number = {2},
  pages = {025005},
  publisher = {American Physical Society},
  doi = {10.1103/RevModPhys.91.025005}
}

@article{forn-diazUltrastrongCouplingSingle2017,
  title = {Ultrastrong Coupling of a Single Artificial Atom to an Electromagnetic Continuum in the Nonperturbative Regime},
  author = {{Forn-D{\'i}az}, P. and {Garc{\'i}a-Ripoll}, J. J. and Peropadre, B. and Orgiazzi, J.-L. and Yurtalan, M. A. and Belyansky, R. and Wilson, C. M. and Lupascu, A.},
  year = 2017,
  month = jan,
  journal = {Nature Phys},
  volume = {13},
  number = {1},
  pages = {39--43},
  publisher = {Nature Publishing Group},
  issn = {1745-2481},
  doi = {10.1038/nphys3905}
}

@article{hennessyQuantumNatureStrongly2007,
  title = {Quantum Nature of a Strongly Coupled Single Quantum Dot--Cavity System},
  author = {Hennessy, K. and Badolato, A. and Winger, M. and Gerace, D. and Atat{\"u}re, M. and Gulde, S. and F{\"a}lt, S. and Hu, E. L. and Imamo{\u g}lu, A.},
  year = 2007,
  month = feb,
  journal = {Nature},
  volume = {445},
  number = {7130},
  pages = {896--899},
  publisher = {Nature Publishing Group},
  issn = {1476-4687},
  doi = {10.1038/nature05586}
}

@article{leibfriedQuantumDynamicsSingle2003,
  title = {Quantum Dynamics of Single Trapped Ions},
  author = {Leibfried, D. and Blatt, R. and Monroe, C. and Wineland, D.},
  year = 2003,
  month = mar,
  journal = {Rev. Mod. Phys.},
  volume = {75},
  number = {1},
  pages = {281--324},
  publisher = {American Physical Society},
  doi = {10.1103/RevModPhys.75.281}
}

@article{niemczykCircuitQuantumElectrodynamics2010,
  title = {Circuit Quantum Electrodynamics in the Ultrastrong-Coupling Regime},
  author = {Niemczyk, T. and Deppe, F. and Huebl, H. and Menzel, E. P. and Hocke, F. and Schwarz, M. J. and {Garcia-Ripoll}, J. J. and Zueco, D. and H{\"u}mmer, T. and Solano, E. and Marx, A. and Gross, R.},
  year = 2010,
  month = oct,
  journal = {Nature Phys},
  volume = {6},
  number = {10},
  pages = {772--776},
  publisher = {Nature Publishing Group},
  issn = {1745-2481},
  doi = {10.1038/nphys1730}
}

@article{schiroPhaseTransitionLight2012a,
  title = {Phase {{Transition}} of {{Light}} in {{Cavity QED Lattices}}},
  author = {Schir{\'o}, M. and Bordyuh, M. and {\"O}ztop, B. and T{\"u}reci, H. E.},
  year = 2012,
  month = aug,
  journal = {Phys. Rev. Lett.},
  volume = {109},
  number = {5},
  pages = {053601},
  publisher = {American Physical Society},
  doi = {10.1103/PhysRevLett.109.053601}
}

@article{yi-xiangGoldstoneHiggsModes2013,
  title = {Goldstone and {{Higgs}} Modes of Photons inside a Cavity},
  author = {Yu, Yixiang and Ye, Jinwu and Liu, Wu-Ming},
  year = 2013,
  month = dec,
  journal = {Sci Rep},
  volume = {3},
  number = {1},
  pages = {3476},
  publisher = {Nature Publishing Group},
  issn = {2045-2322},
  doi = {10.1038/srep03476}
}

@article{erlingssonEnergySpectraQuantum2010,
  title = {Energy Spectra for Quantum Wires and Two-Dimensional Electron Gases in Magnetic Fields with {{Rashba}} and {{Dresselhaus}} Spin-Orbit Interactions},
  author = {Erlingsson, Sigurdur I. and Egues, J. Carlos and Loss, Daniel},
  year = 2010,
  month = oct,
  journal = {Phys. Rev. B},
  volume = {82},
  number = {15},
  pages = {155456},
  issn = {1098-0121, 1550-235X},
  doi = {10.1103/PhysRevB.82.155456}
}

@article{Bin2018PRA_98_043858a,
  title = {Two-Photon Blockade in a Cascaded Cavity-Quantum-Electrodynamics System},
  author = {Bin, Qian and L{\"u}, Xin-You and Bin, Shang-Wu and Wu, Ying},
  year = 2018,
  month = oct,
  journal = {Phys. Rev. A},
  volume = {98},
  number = {4},
  pages = {043858},
  doi = {10.1103/PhysRevA.98.043858},
  langid = {english}
}

@article{Bin2024PRL_133_043601,
  title = {Nonreciprocal {{Bundle Emissions}} of {{Quantum Entangled Pairs}}},
  author = {Bin, Qian and Jing, Hui and Wu, Ying and Nori, Franco and L{\"u}, Xin-You},
  year = 2024,
  month = jul,
  journal = {Phys. Rev. Lett.},
  volume = {133},
  number = {4},
  pages = {043601},
  doi = {10.1103/PhysRevLett.133.043601}
}

@article{Jing2014PRL_113_053604,
  title = {{{PT}} -{{Symmetric Phonon Laser}}},
  author = {Jing, Hui and {\"O}zdemir, S. K. and L{\"u}, Xin-You and Zhang, Jing and Yang, Lan and Nori, Franco},
  year = 2014,
  month = jul,
  journal = {Phys. Rev. Lett.},
  volume = {113},
  number = {5},
  pages = {053604},
  doi = {10.1103/PhysRevLett.113.053604},
  copyright = {http://link.aps.org/licenses/aps-default-license},
  langid = {english}
}

@article{Garziano2013PRA_88_063829,
  title = {Switching on and off of Ultrastrong Light-Matter Interaction: {{Photon}} Statistics of Quantum Vacuum Radiation},
  shorttitle = {Switching on and off of Ultrastrong Light-Matter Interaction},
  author = {Garziano, L. and Ridolfo, A. and Stassi, R. and Di Stefano, O. and Savasta, S.},
  year = 2013,
  month = dec,
  journal = {Phys. Rev. A},
  volume = {88},
  number = {6},
  pages = {063829},
  doi = {10.1103/PhysRevA.88.063829},
  langid = {american}
}

@article{Ridolfo2013PRA_88_063812,
  title = {Photon Correlations from Ultrastrong Optical Nonlinearities},
  author = {Ridolfo, Alessandro and Del Valle, Elena and Hartmann, Michael J.},
  year = 2013,
  month = dec,
  journal = {Phys. Rev. A},
  volume = {88},
  number = {6},
  pages = {063812},
  doi = {10.1103/PhysRevA.88.063812},
  copyright = {http://link.aps.org/licenses/aps-default-license},
  langid = {english}
}

@book{1998__,
  title = {Nobel {{Lectures}} in {{Physics}} 1901 -- 1921},
  year = 1998,
  month = nov,
  doi = {10.1142/3726},
  publisher = {WORLD SCIENTIFIC},
  isbn = {978-981-02-3401-0},
  langid = {english}
}

@book{Carmichael2008__,
  title = {Statistical {{Methods}} in {{Quantum Optics}} 2},
  author = {Carmichael, Howard J.},
  year = 2008,
  series = {Theoretical and {{Mathematical Physics}}},
  address = {Berlin, Heidelberg},
  publisher = {Springer},
  doi = {10.1007/978-3-540-71320-3},
  copyright = {http://www.springer.com/tdm},
  isbn = {978-3-540-71319-7 978-3-540-71320-3}
}

@book{Haroche2006__,
  title = {Exploring the {{Quantum}}: {{Atoms}}, {{Cavities}}, and {{Photons}}},
  shorttitle = {Exploring the {{Quantum}}},
  author = {Haroche, Serge and Raimond, Jean-Michel},
  year = 2006,
  month = aug,
  publisher = {Oxford University Press},
  doi = {10.1093/acprof:oso/9780198509141.001.0001},
  isbn = {978-0-19-850914-1}
}

@article{Lu2023PRA_108_053712,
  title = {$\mathcal{PT}$-Symmetric Quantum {{Rabi}} Model},
  author = {Lu, Xilin and Li, Hui and Shi, Jia-Kai and Fan, Li-Bao and Mangazeev, Vladimir and Li, Zi-Min and Batchelor, Murray T.},
  year = 2023,
  month = nov,
  journal = {Phys. Rev. A},
  volume = {108},
  number = {5},
  pages = {053712},
  doi = {10.1103/PhysRevA.108.053712},
  langid = {english}
}

@article{Wang2025PRA_112_043704,
  title = {Vacuum {{Rabi}} Splitting and Quantum {{Fisher}} Information of a Non-{{Hermitian}} Qubit in a Single-Mode Cavity},
  author = {Wang, Yi-Cheng and Li, Jiong and Duan, Li-Wei and Chen, Qing-Hu},
  year = 2025,
  month = oct,
  journal = {Phys. Rev. A},
  volume = {112},
  number = {4},
  pages = {043704},
  doi = {10.1103/m7kc-vc9h},
  langid = {english}
}

@article{Wang2025AQT_n/a_e00622,
  title = {$\mathcal{PT}$-{{Symmetric Two-Photon Quantum Rabi Models}}},
  author = {Wang, Yi-Cheng and Li, Jiong and Chen, Qing-Hu},
  year = {2025},
  journal = {Adv Quantum Tech},
  volume = {n/a},
  number = {n/a},
  pages = {e00622},
  doi = {10.1002/qute.202500622},
  langid = {english}
}

@article{Tian2023JPBAMOP_56_095001a,
  title = {The Interplay of $\mathcal{PT}$ Symmetry with the {{Rabi}} Oscillation in the Non-{{Hermitian}} Double {{Jaynes}}--{{Cummings}} Model},
  author = {Tian, Zekai and Man, Zhong-Xiao and Zhu, Baogang},
  year = 2023,
  month = apr,
  journal = {J. Phys. B: At. Mol. Opt. Phys.},
  volume = {56},
  number = {9},
  pages = {095001},
  doi = {10.1088/1361-6455/acc776},
  langid = {english}
}

@article{Ying2024AQT_7_2400288,
  title = {Universal {{Quantum Fisher Information}} and {{Simultaneous Occurrence}} of {{Landau-Class}} and {{Topological-Class Transitions}} in {{Non-Hermitian Jaynes-Cummings Models}}},
  author = {Ying, Zu-Jian},
  year = {2024},
  journal = {Adv Quantum Tech},
  volume = {7},
  number = {10},
  pages = {2400288},
  doi = {10.1002/qute.202400288},
  langid = {english}
}

@article{Li2025AQT_8_2400609,
  title = {$\mathcal{PT}$-{{Symmetric Quantum Rabi Model}}: {{Solutions}} and {{Exceptional Points}}},
  shorttitle = {$\mathcal{PT}$-{Symmetric Quantum Rabi Model}},
  author = {Li, Jiong and Wang, Yi-Cheng and Duan, Li-Wei and Chen, Qing-Hu},
  year = {2025},
  journal = {Adv Quantum Tech},
  volume = {8},
  number = {6},
  pages = {2400609},
  doi = {10.1002/qute.202400609},
  langid = {english}
}

@Article{jiang2025cpl,
title = {Quantum and Semiclassical Non-Hermitian Dicke Models without Nonreciprocity},
journal = {Chin. Phys. Lett.},
volume = {42},
number = {12},
pages = {120403},
year = {2025},
issn = {},
doi = {10.1088/0256-307X/42/12/120403},	
url = {http://cpl.iphy.ac.cn/en/article/doi/10.1088/0256-307X/42/12/120403},
author = {Bin Jiang and Yi-Yang Li and Jun-Jie Liu and Chen Wang and Jian-Hua Jiang}
}

@article{chen2022prr,
  title = {Tuning nonequilibrium heat current and two-photon statistics via composite qubit-resonator interaction},
  author = {Chen, Ze-Huan and Che, Han-Xin and Chen, Zhe-Kai and Wang, Chen and Ren, Jie},
  journal = {Phys. Rev. Res.},
  volume = {4},
  issue = {1},
  pages = {013152},
  numpages = {12},
  year = {2022},
  month = {Feb},
  publisher = {American Physical Society},
  doi = {10.1103/PhysRevResearch.4.013152},
  url = {https://link.aps.org/doi/10.1103/PhysRevResearch.4.013152}
}

@article{settineri2018pra,
  title = {Dissipation and thermal noise in hybrid quantum systems in the ultrastrong-coupling regime},
  author = {Settineri, Alessio and Macr\'{\i}, Vincenzo and Ridolfo, Alessandro and Di Stefano, Omar and Kockum, Anton Frisk and Nori, Franco and Savasta, Salvatore},
  journal = {Phys. Rev. A},
  volume = {98},
  issue = {5},
  pages = {053834},
  numpages = {15},
  year = {2018},
  month = {Nov},
  publisher = {American Physical Society},
  doi = {10.1103/PhysRevA.98.053834},
  url = {https://link.aps.org/doi/10.1103/PhysRevA.98.053834}
}

@article{LeBoite2020AQT_3_1900140a,
  title = {Theoretical {{Methods}} for {{Ultrastrong Light}}--{{Matter Interactions}}},
  author = {Le Boit{\'e}, Alexandre},
  year = 2020,
  journal = {Advanced Quantum Technologies},
  volume = {3},
  number = {7},
  pages = {1900140},
  doi = {10.1002/qute.201900140},
  copyright = {\copyright{} 2020 WILEY-VCH Verlag GmbH \& Co. KGaA, Weinheim},
  langid = {english}
}

@article{Flayac2017PRA_96_053810,
  title = {Unconventional Photon Blockade},
  author = {Flayac, H. and Savona, V.},
  year = 2017,
  month = nov,
  journal = {Phys. Rev. A},
  volume = {96},
  number = {5},
  pages = {053810},
  doi = {10.1103/PhysRevA.96.053810},
  copyright = {https://link.aps.org/licenses/aps-default-license},
  langid = {english}
}

@article{Imamoglu1997PRL_79_1467-1470,
  title = {Strongly {{Interacting Photons}} in a {{Nonlinear Cavity}}},
  author = {Imamo{\=g}lu, A. and Schmidt, H. and Woods, G. and Deutsch, M.},
  year = 1997,
  month = aug,
  journal = {Phys. Rev. Lett.},
  volume = {79},
  number = {8},
  pages = {1467--1470},
  doi = {10.1103/PhysRevLett.79.1467}
}

@article{Zhang2018PRA_97_043858,
  title = {Dicke-Model Simulation via Cavity-Assisted {{Raman}} Transitions},
  author = {Zhang, Zhiqiang and Lee, Chern Hui and Kumar, Ravi and Arnold, K. J. and Masson, Stuart J. and Grimsmo, A. L. and Parkins, A. S. and Barrett, M. D.},
  year = 2018,
  month = apr,
  journal = {Phys. Rev. A},
  volume = {97},
  number = {4},
  pages = {043858},
  doi = {10.1103/PhysRevA.97.043858},
  langid = {english}
}

@article{lvQuantumSimulationQuantum2018,
  title = {Quantum {{Simulation}} of the {{Quantum Rabi Model}} in a {{Trapped Ion}}},
  author = {Lv, Dingshun and An, Shuoming and Liu, Zhenyu and Zhang, Jing-Ning and Pedernales, Julen S. and Lamata, Lucas and Solano, Enrique and Kim, Kihwan},
  year = 2018,
  month = apr,
  journal = {Phys. Rev. X},
  volume = {8},
  number = {2},
  pages = {021027},
  publisher = {American Physical Society},
  doi = {10.1103/PhysRevX.8.021027}
}

@article{baksicControllingDiscreteContinuous2014,
  title = {Controlling {{Discrete}} and {{Continuous Symmetries}} in ``{{Superradiant}}'' {{Phase Transitions}} with {{Circuit QED Systems}}},
  author = {Baksic, Alexandre and Ciuti, Cristiano},
  year = 2014,
  month = apr,
  journal = {Phys. Rev. Lett.},
  volume = {112},
  number = {17},
  pages = {173601},
  publisher = {American Physical Society},
  doi = {10.1103/PhysRevLett.112.173601},
  urldate = {2026-02-18}
}

@article{wangSimulatingAnisotropicQuantum2019a,
  title = {Simulating {{Anisotropic}} Quantum {{Rabi}} Model via Frequency Modulation},
  author = {Wang, Gangcheng and Xiao, Ruoqi and Shen, H. Z. and Sun, Chunfang and Xue, Kang},
  year = 2019,
  month = mar,
  journal = {Sci Rep},
  volume = {9},
  number = {1},
  pages = {4569},
  publisher = {Nature Publishing Group},
  issn = {2045-2322},
  doi = {10.1038/s41598-019-40899-7},
  urldate = {2026-02-18},
  copyright = {2019 The Author(s)},
  langid = {english}
}

@book{carmichaelStatisticalMethodsQuantum2008,
  title = {Statistical {{Methods}} in {{Quantum Optics}} 2},
  author = {Carmichael, Howard J.},
  year = 2008,
  series = {Theoretical and {{Mathematical Physics}}},
  publisher = {Springer},
  address = {Berlin, Heidelberg},
  doi = {10.1007/978-3-540-71320-3},
  urldate = {2026-01-15},
  copyright = {http://www.springer.com/tdm},
  isbn = {978-3-540-71319-7 978-3-540-71320-3},
  langid = {american}
}

@article{Yoshihara2017NP_13_44-47,
  title = {Superconducting Qubit-Oscillator Circuit beyond the Ultrastrong-Coupling Regime},
  author = {Yoshihara, Fumiki and Fuse, Tomoko and Ashhab, Sahel and Kakuyanagi, Kosuke and Saito, Shiro and Semba, Kouichi},
  year = 2017,
  month = jan,
  journal = {Nature Physics},
  volume = {13},
  number = {1},
  pages = {44--47},
  doi = {10.1038/nphys3906},
  langid = {english}
}

@article{Askenazi2014NJP_16_043029,
  title = {Ultra-Strong Light--Matter Coupling for Designer {{Reststrahlen}} Band},
  author = {Askenazi, B and Vasanelli, A and Delteil, A and Todorov, Y and Andreani, L C and Beaudoin, G and Sagnes, I and Sirtori, C},
  year = 2014,
  month = apr,
  journal = {New J. Phys.},
  volume = {16},
  number = {4},
  pages = {043029},
  doi = {10.1088/1367-2630/16/4/043029},
  langid = {english}
}

\end{document}